# Scan Specific Artifact Reduction in K-space (SPARK) Neural Networks Synergize with Physics-based Reconstruction to Accelerate MRI

Yamin Arefeen[1], Onur Beker[2], Jaejin Cho[3], Heng Yu[4], Elfar Adalsteinsson[1,5,6], Berkin Bilgic[3,5,7]

1 Department of Electrical Engineering and Computer Science, Massachusetts Institute of Technology, Cambridge, MA, USA
2 Computer and Communication Sciences, École Polytechnique Fédérale de Lausanne, Lausanne, Switzerland
3 Athinoula A. Martinos Center for Biomedical Imaging, Charlestown, MA, USA
4 Department of Automation, Tsinghua University, Beijing, China
5 Harvard-MIT Health Sciences and Technology, Massachusetts Institute of Technology, Cambridge, MA, USA
6 Institute for Medical Engineering and Science, Massachusetts Institute of Technology, Cambridge, MA, USA
7 Department of Radiology, Harvard Medical School, Boston, MA, USA

**Corresponding author:**
Yamin Arefeen
yarefeen@mit.edu

# ABSTRACT

**Purpose:** To develop a scan-specific model that estimates and corrects k-space errors made when reconstructing accelerated Magnetic Resonance Imaging (MRI) data.

**Methods:** Scan-Specific Artifact Reduction in k-space (SPARK) trains a convolutional-neural-network to estimate and correct k-space errors made by an input reconstruction technique by back-propagating from the mean-squared-error loss between an auto-calibration signal (ACS) and the input technique's reconstructed ACS. First, SPARK is applied to GRAPPA and demonstrates improved robustness over other scan-specific models, such as RAKI and residual-RAKI. Subsequent experiments demonstrate that SPARK synergizes with residual-RAKI to improve reconstruction performance. SPARK also improves reconstruction quality when applied to advanced acquisition and reconstruction techniques like 2D virtual coil (VC-) GRAPPA, 2D LORAKS, 3D GRAPPA without an integrated ACS region, and 2D/3D wave-encoded imaging.

**Results:** SPARK yields SSIM improvement and 1.5 - 2x RMSE reduction when applied to GRAPPA and improves robustness to ACS size for various acceleration rates in comparison to other scan-specific techniques. When applied to advanced reconstruction techniques such as residual-RAKI, 2D VC-GRAPPA and LORAKS, SPARK achieves up to 20% RMSE improvement. SPARK with 3D GRAPPA also improves RMSE performance by ~2x, SSIM performance, and perceived image quality without a fully sampled ACS region. Finally, SPARK synergizes with non-cartesian, 2D and 3D wave-encoding imaging by reducing RMSE between 20 - 25% and providing qualitative improvements.

**Conclusion:** SPARK synergizes with physics-based acquisition and reconstruction techniques to improve accelerated MRI by training scan-specific models to estimate and correct reconstruction errors in k-space.

# INTRODUCTION

Magnetic Resonance Imaging (MRI) acquisitions suffer from an inherent tradeoff between scan time, resolution, and signal-to-noise ratio (SNR).[1] Reducing scan time while maintaining image quality has been a vibrant area of research over the last couple of decades as accelerated MRI enables improved patient comfort[2], superior clinical efficiency[3], and decreased power deposition into the body[4].

Clinical scans routinely employ parallel imaging[5–7] to accelerate MRI acquisitions. These techniques omit k-space measurements and interpolate missing data using the encoding capabilities of modern multi-coil receive arrays. Generalized autocalibrating partially parallel acquisitions (GRAPPA) applies linear convolutional kernels, trained on a fully sampled auto-calibration signal (ACS) region, to synthesize missing k-space lines. Sensitivity encoding (SENSE) utilizes explicit knowledge of receive profiles to recover an image from undersampled data through an inverse problem formulation[8]. While widespread, these linear techniques suffer from increased noise amplification due to ill-conditioning of the reconstruction problem at higher acceleration rates. To improve parallel imaging, methods such as Iterative self-consistent Parallel Imaging Reconstruction (SPIRiT)[9] and Low-rank modeling of local k-space neighborhoods (LORAKS)[10], apply sophisticated priors on the structure and redundancy in k-space data. GRAPPA also can be improved through k-space specific Tikhonov regularization based on the local noise levels of the ACS region[11]. On the acquisition side, CAIPIRINHA for 3D imaging and blipped-CAIPI for simultaneous-multislice (SMS) improves conditioning in cartesian parallel imaging through tailored aliasing[12–14]. Similarly, wave-encoding employs non-cartesian corkscrew trajectories to better pose parallel imaging inversion by spreading aliasing along all spatial dimensions[15].

Tailored undersampling and non-linear reconstructions that impose priors through regularization provide an alternative approach to reducing MRI scan time. In compressed sensing,[16] undersampled data are acquired randomly. The resultant incoherent aliasing artifacts can be mitigated with l1-regularization[16] and low-rank priors[17,18] imposed in a variety of domains including gradient[19], wavelet[2], explicit low-rank coefficients[20], and learned dictionaries[21]. Additionally, compressed sensing can be combined with parallel imaging[22], with substantial benefits in high dimensional imaging.[20,21,23,24] Limited achievable incoherence in 2D imaging[16] precludes widespread use of compressed sensing in standard slice-by-slice acquisitions.

Machine learning provides yet another alternative for accelerating MRI. Models, typically neural networks trained in a supervised fashion, apply regularization, directly reconstruct the undersampled data, and correct image-domain reconstruction artifacts[25–32]. Standard supervised learning achieves impressive results, but requires large amounts of training data and may not generalize reliably outside of the particular application upon which they were trained[33] .

To address generalizability and data availability, scan-specific robust artificial-neural-networks for k-space interpolation (RAKI)[34] elegantly proposes training a neural network on just the ACS of a specific scan to estimate undersampled lines in k-space. RAKI enables subject specific accelerated MRI without vast amounts of training data. In addition, it functions with uniform, non-uniform,[35] and 3D undersampling, regularized reconstructions, and non-linear autoregressive modeling.[36,37] Despite limited ACS size, RAKI yields remarkable performance improvements for accelerated MRI.

We propose an alternative k-space based, scan specific model for improving accelerated MRI reconstruction. Rather than interpolating missing k-space data, our method takes a reconstruction of the undersampled data using any technique as input and estimates the k-space reconstruction artifacts in the input. The input reconstruction can be k-space based, like GRAPPA, or can be model-based, like SENSE. Model-based reconstructed k-space can be generated by applying the physics based forward model governing data acquisition to the model-based reconstruction. We can perform Scan Specific Artifact Reduction in k-space (SPARK) by taking our estimate of the residual and subtracting it from the reconstructed estimate of k-space. SPARK can be seen as scan-specific refinement that flexibly applies to arbitrary acquisition and reconstruction schemes.

Our technique bears resemblance to residual-RAKI[38], but residual-RAKI only estimates residuals for 2D-cartesian, under-sampled k-space reconstructed with a linear convolutional layer of a neural network. On the other hand, SPARK flexibly corrects artifacts in a wide range of physics-based acquisition and reconstruction schemes beyond just the 2D cartesian, linear reconstruction setting. Additionally, SPARK can synergize with and be applied to improve residual-RAKI reconstructions, as demonstrated herein.

We begin by describing how SPARK trains a non-linear model to estimate and correct k-space residuals of an input reconstruction using scan-specific ACS data. Next, we compare GRAPPA, SPARK, RAKI, and residual-RAKI with 1D acceleration and demonstrate that correcting a GRAPPA reconstruction with SPARK yields more robust reconstruction improvements at lower ACS sizes in comparison to RAKI and residual-RAKI. Additionally, we demonstrate that SPARK synergizes with residual-RAKI by applying SPARK to improve residual-RAKI reconstructions. We also show that SPARK integrates with more advanced parallel-imaging reconstruction techniques like virtual-coil (VC) GRAPPA[39] and LORAKS and compare the results to residual-RAKI. Then, we extend SPARK to volumetric imaging by comparing SPARK to standard GRAPPA in 3D cartesian reconstructions with an integrated ACS region in the acquisition. Second, a framework to apply SPARK without a fully sampled integrated ACS region is proposed and demonstrated. Finally, we apply SPARK in single-slice and 3D non-cartesian wave-encoded imaging, a setting in which standard RAKI and residual-RAKI cannot be applied in their current form.

Initial versions of this work were presented as abstracts at ISMRM[40,41].

## METHODS

### SPARK Procedure: Estimating and Correcting Reconstruction Errors in K-space

Let $x_{est} \in \mathbb{C}^{M \times N \times P \times C}$ and $y_{est} \in \mathbb{C}^{N_{ro} \times N_{pe} \times N_{pa} \times C}$ be the reconstructed multi-coil image and k-space using some reconstruction technique given acquired data $y_{acq} \in \mathbb{C}^{N_{ro} \times N_{pe} \times N_{pa} \times C}$, where $M \times N \times P$ are image dimensions, $C$ is the number of coils, and $N_{ro} \times N_{pe} \times N_{pa}$ are k-space dimensions. Let $A$ be the operator which projects k-space data onto the ACS region.

Separate variables represent k-space and image dimensions because acquired and reconstructed k-space and image dimensions can differ in the wave-encoding applications explored later. Subsequent 2D cartesian experiments use dimensions $(N_{ro} \times N_{pe} \times N_{pa} = M \times N \times P = 236 \times 188 \times 1)$, 3D cartesian experiments use $(N_{ro} \times N_{pe} \times N_{pa} = M \times N \times P = 186 \times 168 \times 160)$, and 2D wave-encoded experiments use $(N_{ro} \times N_{pe} \times N_{pa} = 768 \times 256 \times 1$ and $M \times N \times P = 256 \times 256 \times 1)$.

We pose the following optimization problem to train a model, $f_{\theta_c}^c$, which estimates reconstruction error in the $c^{th}$ coil, by modifying model parameters $\theta_c$.

$$\theta_c^* = argmin_\theta \left\| A[y_{acq}^c - y_{est}^c - f_{\theta_c}^c(y_{est})] \right\|_2 \qquad \text{Eq1}$$

In equation 1, we determine model parameters, $\theta_c^*$, with which $f_{\theta_c^*}^c$ estimates the k-space reconstruction error inside of the ACS for the $c^{th}$ coil given the reconstructed k-space across all coils as input. Since SPARK only has access to the measured and reconstructed data, each $f_{\theta_c}^c$ will be trained with one input and output pair. The input to $f_{\theta_c}^c$ will be the multi-coil reconstructed k-space, $y_{est}$, corresponding to a tensor of dimension $\mathbb{C}^{N_{ro} \times N_{pe} \times N_{pa} \times C}$. $f_{\theta_c}^c$ outputs an estimated k-space correction term for coil $c$, $f_{\theta_c}^c(y_{est})$ of dimension $\mathbb{C}^{N_{ro} \times N_{pe} \times N_{pa} \times 1}$. Note, tensors are not slid around to generate additional samples. To compute training loss, the estimated correction, $f_{\theta_c}^c(y_{est})$, is cropped to the ACS region, and then compared to the difference between the measured and reconstructed ACS region for the k-space of coil $c$, denoted as $A[y_{acq}^c - y_{est}^c]$.

Since these models only need to estimate errors in the scan to be reconstructed, and not a wide range of scans, one input / output pair provides enough information for the model to learn how to estimate the reconstruction errors inside of the calibration region.

The trained model then corrects reconstruction error in the estimated k-space of the $c^{th}$ coil:

$$y_{corrected}^c = y_{est}^c + f_{\theta_c^*}^c(y_{est}) \qquad \text{Eq2}$$

Although SPARK trains models to estimate errors inside ACS, equation 2 applies the correction estimated by $f_{\theta_c^*}^c$ to the entire k-space of the $c^{th}$ coil. Training a model and applying a correction with Eq1 and Eq2 is repeated for each coil, until all estimated k-space has been corrected. Then, a fully sampled reconstruction (typically involving an inverse Fourier transform and coil-combination) generates the reconstructed image from the corrected k-space. **Fig 1** illustrates SPARK applied to a cartesian GRAPPA reconstruction, where the k-space and image space matrix sizes are equivalent.

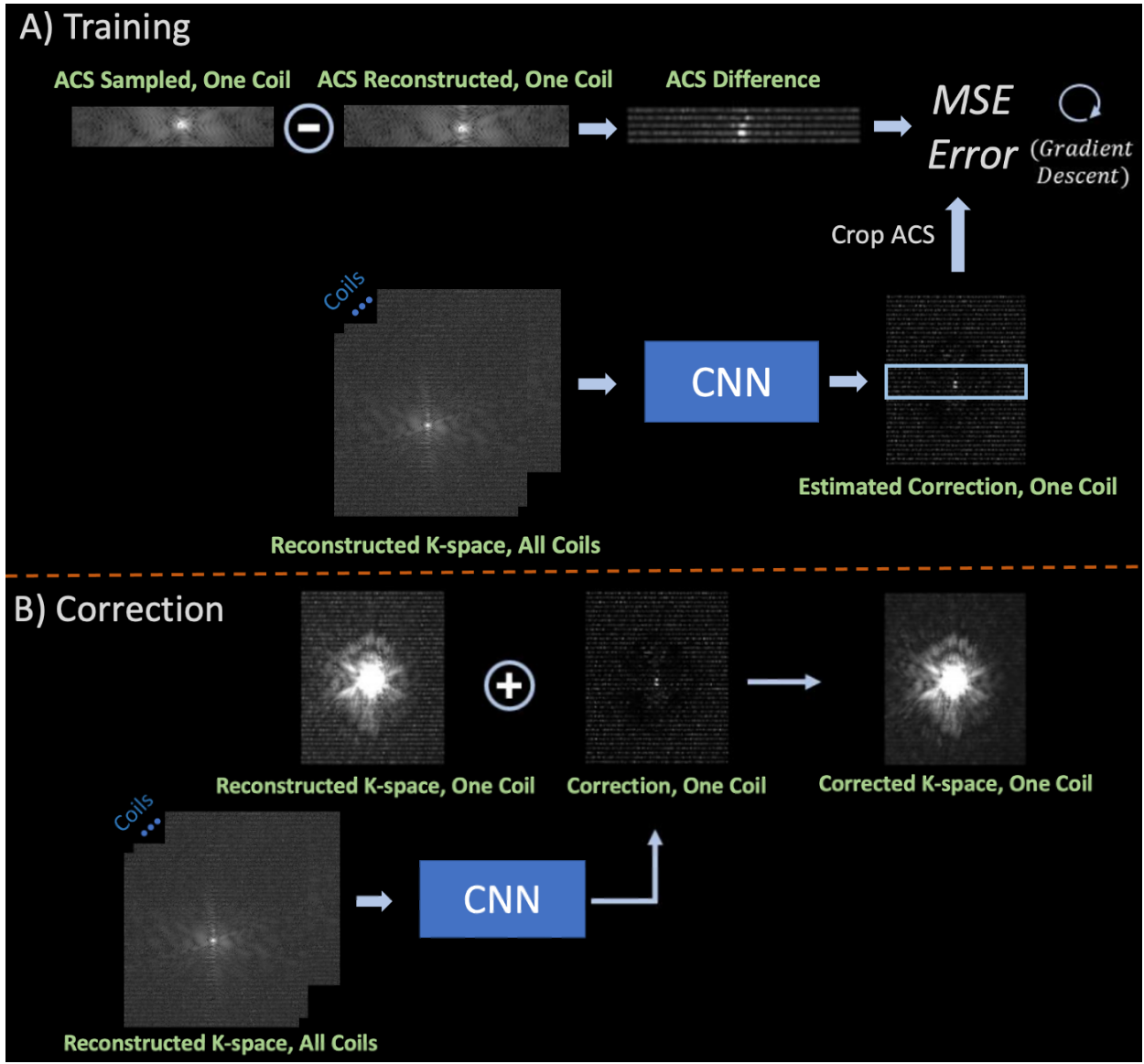

*Fig 1.* A schematic overview of the SPARK procedure using a GRAPPA reconstruction as input. In this exemplary cartesian setting, the k-space and image dimensions are equivalent ($N_{ro} \times N_{pe} \times N_{pa} = M \times N \times P = 236 \times 188 \times 1$). *(A)* We subtract the measured and reconstructed ACS data for the $c^{th}$ coil to compute an ACS difference. Then, we pass the input reconstructed k-space across all coils to estimate a correction term. The correction term is cropped and compared to the ACS difference through least-squares loss, and backpropagation can be applied to adjust model parameters. *(B)* To apply corrections, the trained model will estimate a correction term using all reconstructed k-space as input and add the resultant correction to the $c^{th}$ coil. By repeating this process for all *C* coils, k-space artifacts can be corrected in a scan-specific manner for a wide range of reconstruction and acquisition settings.

Like GRAPPA and RAKI, SPARK assumes models trained to estimate reconstruction artifacts inside the ACS will generalize and estimate errors outside of the region. Subsequent experiments will show empirically that SPARK effectively applies corrections to external k-space regions. Additionally, **Supporting Figure 1** and **Supporting Figure** 2 demonstrate SPARK's ability to perform k-space correction outside of the ACS region in cartesian and wave-encoded settings.

Our proposed framework differs from residual-RAKI which attempts to simultaneously interpolate the missing k-space data with a learned estimate from a linear convolutional layer and estimate a correction term of the learned linear reconstruction.[38] Rather, SPARK utilizes any prior reconstruction of the under-sampled k-space as input, enabling integration with a wide range of reconstruction and acquisition techniques, including residual-RAKI.

## Choice of Model for Residual Estimation

Like other applications of machine learning for accelerated MRI, we choose a convolutional neural network (CNN) model for $f_\theta^c$.[42] CNNs demonstrate remarkable performance in approximating nonlinear functions in a wide range of tasks[43] and estimating image-based residuals through supervised learning in a variety of MRI imaging settings.[29–32,44] As elegantly demonstrated by RAKI[34], CNNs can interpolate missing k-space lines in multi-coil acquisitions using just ACS data from the specific scan for training.

We utilize the ADAM optimizer[45] with a fixed learning rate for training. To operate on complex data, we concatenate the real and imaginary portions of k-space along the channel dimension of CNN inputs and outputs.[42] All models were implemented with Python and PyTorch[46] and employed no bias terms in its layers, as they may cause issues with k-space scaling.[34] Detailed network structure and applied non-linearity varies across experiments and will be described in subsequent sections.

## In-vivo Single-Slice Cartesian Experiments

*Imaging Experiments*

All imaging protocols were performed with approval from the local institutional review board with written informed consent. A volunteer was scanned with a Siemens 3T Magnetom Skyra (Siemens Healthineers, Erlangen, Germany) system using a 32-channel receive head coil. A fully sampled MPRAGE scan was acquired with: FOV = 234 x 188 x 192 mm$^3$, voxel size = 1 x 1 x 1 mm$^3$, TR/TE/TI = 2530/1.7/1100 ms, flip angle = 7°, bandwidth = 651 Hz/pixel. The raw data were Fourier transformed along the slice (third) dimension, and a single axial slice was used for subsequent analysis (234 x 188 matrix size).

The data were retrospectively undersampled in the AP direction with $R_{in\text{-}plane}$ = {5, 6} and {20, 24, 30, 36, 40} ACS lines. SPARK was compared to GRAPPA with optimized kernel sizes, Tikhonov regularization, and ACS replacement, RAKI, and residual-RAKI.[34,47] SPARK took GRAPPA reconstructions *without* ACS replacement as input.

Additionally, the same retrospectively under-sampled dataset was used to demonstrate synergy between residual-RAKI and SPARK. SPARK corrected k-space errors in the residual-RAKI reconstructions at $R_{in-plane}$ = {5, 6} and with {24, 30, 36, 40, 44} ACS lines.

Next, the single-slice, MPRAGE data was retrospectively undersampled at $R_{in-plane}$ = {5,6} in the AP direction with 30 ACS lines. Data were reconstructed with VC GRAPPA[39,48] and LORAKS[10] and used as input to SPARK. For the LORAKS input, we estimated LORAKS subspaces from the ACS, undersampled the ACS, and used the reconstruction *without* ACS replacement as input. We compared SPARK to VC GRAPPA and LORAKS reconstructions with the ACS region included and to residual-RAKI to illustrate the benefits of applying SPARK to advanced reconstruction techniques.

To evaluate SPARK on inputs with differing noise levels but similar residual artifacts, the $R_{in-plane}$ = {5,6}, MPRAGE data were reconstructed with GRAPPA and VC-GRAPPA and corrected with SPARK using 30 ACS lines. Additionally, we applied SPARK corrections to the $R_{in-plane}$ = 5 data with 30 ACS lines reconstructed using GRAPPA with varying Tikhonov regularization values [0, 1, 5, 10, 50]. Tikhonov regularization was applied to k-space with maximum absolute value of 14.44 and L2-norm of 109.67.

Finally, we averaged 10 slices from the MPRAGE volume, yielding a 1 x 1 x 10 mm axial slice, to perform experiments in a high SNR regime. First, the data were retrospectively under-sampled with $R_{in-plane}$ = {6,7}, reconstructed with GRAPPA and LORAKS, and corrected with SPARK using 30 ACS lines. Second, the high SNR data were retrospectively under-sampled in the AP direction with $R_{in-plane}$ = {6, 7} and with {24, 30, 36, 40, 44} ACS lines. SPARK applied to GRAPPA was compared to GRAPPA, RAKI, and residual-RAKI.

*Montecarlo-esque Noise Stability Evaluation*

Since SPARK introduces non-linearity into the reconstruction pipeline, closed-form g-factor analysis is not applicable. To evaluate noise performance when applying SPARK to GRAPPA, we applied a version of the pseudo-replica technique[49] on the single-slice MPRAGE data. First, the noise-covariance matrix was estimated, and 100 instances of synthesized k-space was generated by adding gaussian noise correlated across the 32 receive channels to the fully sampled k-space. Next, the original and synthesized k-spaces were under-sampled by $R_{in-plane}$ = {4,5,6} in the AP direction. Using kernels computed from the 30 central lines of the original dataset, we reconstructed the original undersampled k-space and the synthesized undersampled k-spaces with GRAPPA. The SPARK correction model is then trained using the same 30 lines and the GRAPPA reconstruction from the original k-space as input. We apply SPARK corrections to the 100 GRAPPA reconstructions from synthesized k-spaces and combine the multi-coil reconstructed data for both GRAPPA and GRAPPA with SPARK through ESPIRiT complex coil-combination.[50] Finally, we take the standard deviation across the real portion of the 100 GRAPPA and GRAPPA with SPARK reconstructions and divide the GRAPPA and GRAPPA with SPARK reconstructions on the original undersampled k-space respectively by these standard deviations to arrive at a proxy for retained SNR.

All single-slice SPARK experiments described in this section utilized the CNN structure in **Fig 2(a),** consisting of 6 convolution layers with 3 x 3 kernels, ReLU non-linearities[51], and a skip-connection between the input and the output of the third[52] with 200 iterations and a learning rate of .0075. The model takes the estimated real and imaginary portions of k-space across all coils as input and outputs the correction for either the real or imaginary portion of k-space for the current coil to be corrected. Supporting Figure 9 explains our hyperparameter and model choice in detail.

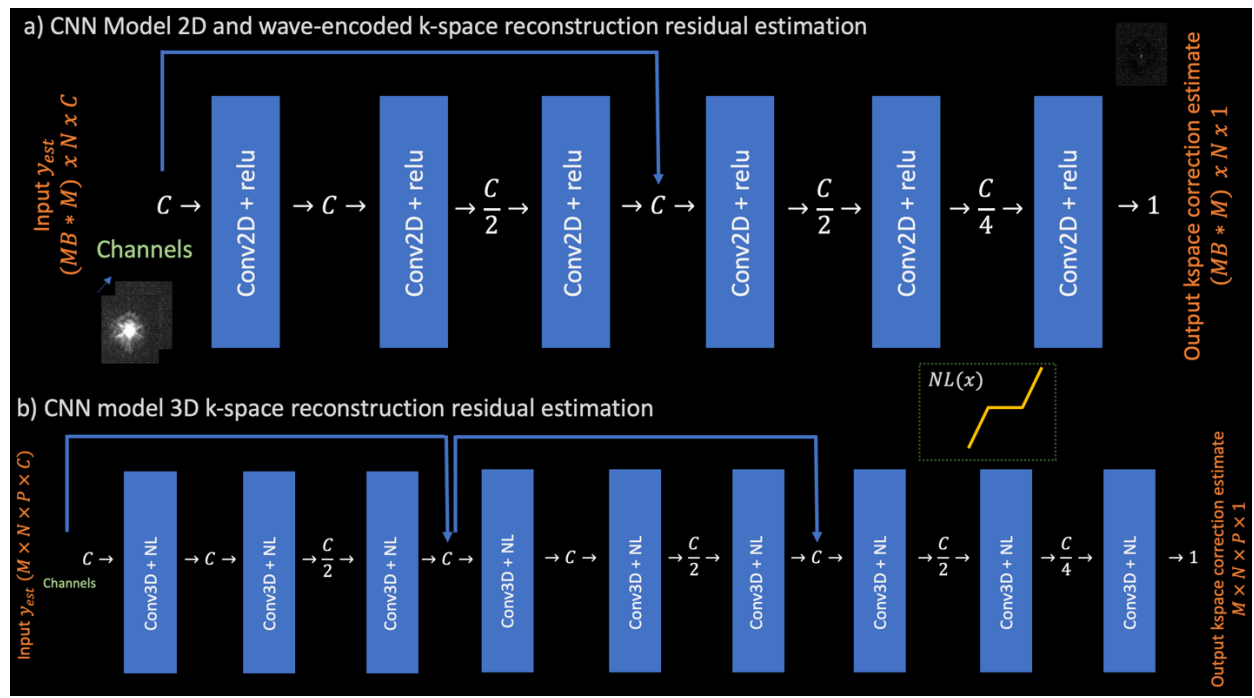

*Fig 2.* Illustrations of the CNN models used in the different experiments. *(A)* For single-slice and wave-encoded experiments, we apply a model with 6 convolutional layers, ReLU non-linearities, 3 x 3 convolutions, and a skip connection between the input and third layer. *(B)* For volumetric experiments, we apply a model with 9 convolutional layers, ReLU non-linearities, 3 x 3 x 3 convolutions, and two skip connections. Due to limited ACS data, models are constrained to be relatively shallow. Models take reconstructed k-space across all coils as input and estimate the k-space reconstruction error for one particular coil.

### In-vivo Volumetric Cartesian Experiments

A second healthy volunteer was scanned with a Siemens 3T Magnetom Skyra (Siemens Healthcare, Erlangen, Germany) system using a 32-channel receive head coil. A fully sampled 3D GRE scan was acquired with: FOV = 244 x 185 x 160 mm$^3$, voxel size = 1.3 x 1.3 x 1 mm$^3$, TR/TE = 25/4 ms, flip angle = 4°, bandwidth = 150 Hz/pixel.

First, we illustrated the feasibility of applying SPARK to volumetric reconstructions and the potential benefit of utilizing an entire 3D block of ACS for model fitting. We retrospectively under-sampled the data by a factor of 4 in the phase encode (second) dimension, and 3 in the partition (third dimension) with a fully sampled 186 x 30 x 30 ACS region. SPARK was compared to 3D-GRAPPA with ACS replacement. SPARK took the corresponding 3D-GRAPPA reconstruction *without* ACS replacement as input.

Second, it can be more efficient to exclude integrated ACS during contrast encoding, and instead utilize calibration scans. Thus, we acquired a minimum TE/TR GRE reference (64 x 32 x 32 matrix size). We propose a hybrid sampling scheme where the ACS region and outer portions of k-space are undersampled at different rates, as suggested in the original RAKI paper.[34] For SPARK, we undersampled the 186 x 48 x 48 integrated ACS region by a factor of 3 x 2 in the phase and partition encoding dimensions and the exterior of k-space by a factor of 4 x 3 with elliptical undersampling,[53] yielding a net acceleration factor of 12. GRAPPA kernels estimated from the reference reconstructed the undersampled 186 x 48 x 48 ACS region. Then, kernels estimated from the reconstructed ACS perform 4 x 3 accelerated GRAPPA reconstruction without ACS replacement, which serves as input to SPARK. SPARK+GRAPPA was compared to 3D GRAPPA on 4 x 3 retrospectively undersampled data with no ACS region and kernels trained on the reference, also yielding a net acceleration factor of 12. We performed experiments without regularization in any reconstruction and with an optimized Tikhonov regularization parameter for both the GRAPPA input to SPARK and 3D GRAPPA baseline.

All 3D SPARK experiments utilize the CNN structure illustrated in **Fig 2(b)**, which consists of 9 convolution layers with 3 x 3 x 3 kernels and skip-connections between the input and third layer output and the third layer output and sixth layer output. We take advantage of the larger 3D ACS region by increasing the complexity of the model, in comparison to the 2D regime. **Supporting Text** 1 provides further discussion on hyper-parameter and model choice in the 3D setting. Finally, applying the following custom nonlinearity improved 3D results in comparison to the ReLU nonlinearity, which produced structured biases in the reconstructed images[54]:

$$NL(x) = x + ReLU\left(\frac{x-1}{2}\right) + ReLU\left(-\frac{x+1}{2}\right) \quad\quad Eq3$$

### In-vivo Wave-encoded Experiments

In-vivo wave-encoded MPRAGE[55] data were acquired with IRB approval using a Siemens 3T Magnetom Skyra (Siemens Healthcare, Erlangen, Germany) system and 32-channel receive head coil. The dataset was fully phase encoded and oversampled by a factor of 3x along the readout dimension to accommodate wave-encoding.[15] The following imaging parameters were utilized: nominal FOV: 256 x 256 x 192 mm$^3$, voxel size = 1 x 1 x 1 mm$^3$, TR/TE/TE = 2500/3.48/1100 ms, flip angle = 8°, readout duration = 5.08 ms, maximum slew rate of wave gradients = 175 mT/m/s, maximum wave gradient amplitude = 9.4 mT/m, 15 sinusoidal wave cycles, k-space dimensions: 768 x 256 x 256 x 32. The wave point-spread function was calibrated with an auto-calibrated approach.[55,56]

We first demonstrate SPARK's ability to synergize with wave-encoded acquisitions by generating retrospectively undersampled, 2D-wave-encoded data. We applied an inverse fourier transform along the phase and partition dimensions of the raw k-space, and a slice along the partition axis was selected for subsequent analysis. A forward fourier transform along the phase-encode dimension of the slice generated single-slice wave-encoded k-space data. The point-spread-function (PSF) corresponding to the selected slice was used in the wave-encoded forward model. We undersampled the wave encoded data by $R_{in\text{-}plane}$ = {5,6} with a 24 phase encode ACS region (matrix 768 x 24) and apply SPARK to the standard least squares reconstruction with the wave-encoded forward model that uses Fourier encoding, coil sensitivities and wave PSFs to map an image to acquired data.[57] To create an input for SPARK, we estimate $x_{est}$ using the generalized sense forward model excluding ACS data. Then, we can generate $y_{est}$ by taking $x_{est}$ through the forward model and apply the standard SPARK procedure.

Next, we applied SPARK to wave-encoded slice-groups. Volumetric wave-encoded data was undersampled by $R_{phase\text{-}encode}$ = 5 and $R_{partition}$ = 3 with CAIPI sampling[13] and a 768 x 30 ACS region at each sampled partition. Since uniform undersampling in the partition dimension allows reconstruction of each individual slice-group[58], we apply least-squares reconstruction with the forward model for the collapsed slice-group as a baseline. A similar least-squares reconstruction excluding ACS generates an estimate of the collapsed wave-encoded k-space for the slice-group used as input for SPARK. The acquired and reconstructed ACS region in the collapsed k-space is used to train the SPARK models which correct our estimate of the collapsed k-space for the slice-group. A final fully-sampled least squares reconstruction creates corrected images for the slice-group.

Wave-encoded SPARK utilizes the same network structure from the 2D GRAPPA experiments shown in **Fig 2 (a)**. All coil sensitivity maps were estimated with the ESPIRiT algorithm.[50,59]

## Results

### In-vivo Single-Slice Cartesian Imaging Results

**Fig 3** compares GRAPPA, RAKI, residual-RAKI, and SPARK with GRAPPA in in-vivo single-slice MPRAGE experiments. (A) shows reconstructions and error maps with $R_{in\text{-}plane}$ = 5 and 30 ACS lines. RAKI, residual-RAKI, and SPARK with GRAPPA improve RMSE and SSIM in comparison to GRAPPA and reduce image degradation. In (B), two plots compare reconstruction RMSE of GRAPPA, RAKI, residual-RAKI, and SPARK with GRAPPA at $R_{in\text{-}plane}$ = {5,6} and varying ACS sizes. SPARK with GRAPPA performs as well as or better than RAKI and residual-RAKI (and always outperforms GRAPPA), particularly at smaller ACS sizes. (For an example with SSIM, at $R_{in\text{-}plane}$ = 6 and 24 ACS lines: SPARK = 0.8859, RAKI = 0.8214, residual-RAKI RMSE = 0.8481, GRAPPA = 0.7714).

Average computation times for GRAPPA+SPARK, RAKI, and residual-RAKI are 64.7 +/- 2.1 seconds, 246.7 +/- 19.6 seconds, and 553.6 +/- 49.7 seconds respectively with more detailed compute comparisons in **Supporting Figure 10**.

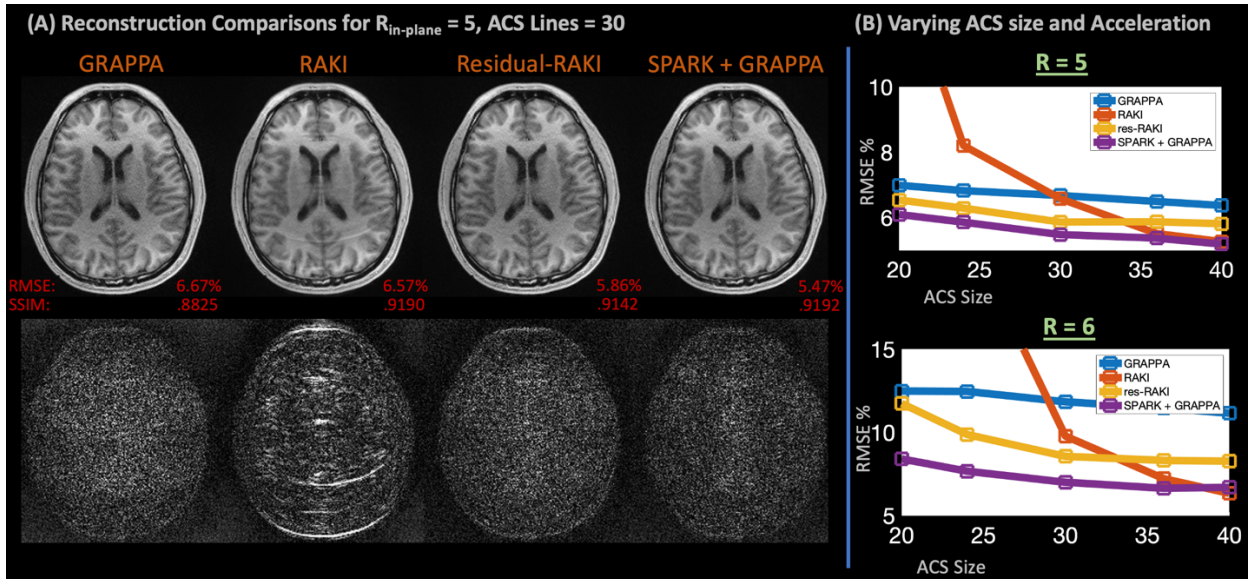

*Fig 3.* (A) Comparison between GRAPPA, RAKI, residual-RAKI, and SPARK applied to an initial GRAPPA reconstruction at $R_{in-plane}$ = 5 and 30 ACS lines. RAKI, residual-RAKI, and SPARK all achieve quantitative RMSE, SSIM, and visual improvements in comparison to GRAPPA. *(B)* Comparisons between GRAPPA, RAKI, residual-RAKI, and SPARK for $R_{in-plane}$ = {5,6} and a range of ACS sizes. SPARK always performs at least as well or better than RAKI and residual-RAKI, particularly at smaller ACS sizes.

**Supporting Figure 3** presents the same experiments from Fig 3 on the higher SNR MPRAGE slice. At $R_{in-plane}$ = 6 with 36 ACS lines, RAKI residual-RAKI, and SPARK with GRAPPA improve RMSE and SSIM in comparison to GRAPPA. The plots of RMSE at $R_{in-plane}$ = {6,7} also suggest that SPARK outperforms RAKI and residual-RAKI at the varying ACS sizes, particularly for smaller calibration regions.

**Fig 4** demonstrates SPARK's ability to synergize with residual-RAKI. With 36 ACS lines and $R_{in-plane}$ = {5,6}, applying SPARK to the residual-RAKI reconstruction improves RMSE, SSIM, and qualitative reconstruction quality, as evidenced by the error maps. Additionally, SPARK improves the RMSE, shown in the plots of Fig. 4, and SSIM of residual-RAKI at a variety of acceleration rates and ACS sizes. Residual-RAKI at $R_{in-plane}$ = 6 with ACS lines of {24,30,36,40} yields SSIM values of {0.748,0.784,0.796,0.804} while residual-RAKI+SPARK achieves values of {0.814,0.843,0.873,0.872}. Similar SSIM improvements can be seen at $R_{in-plane}$ = 5.

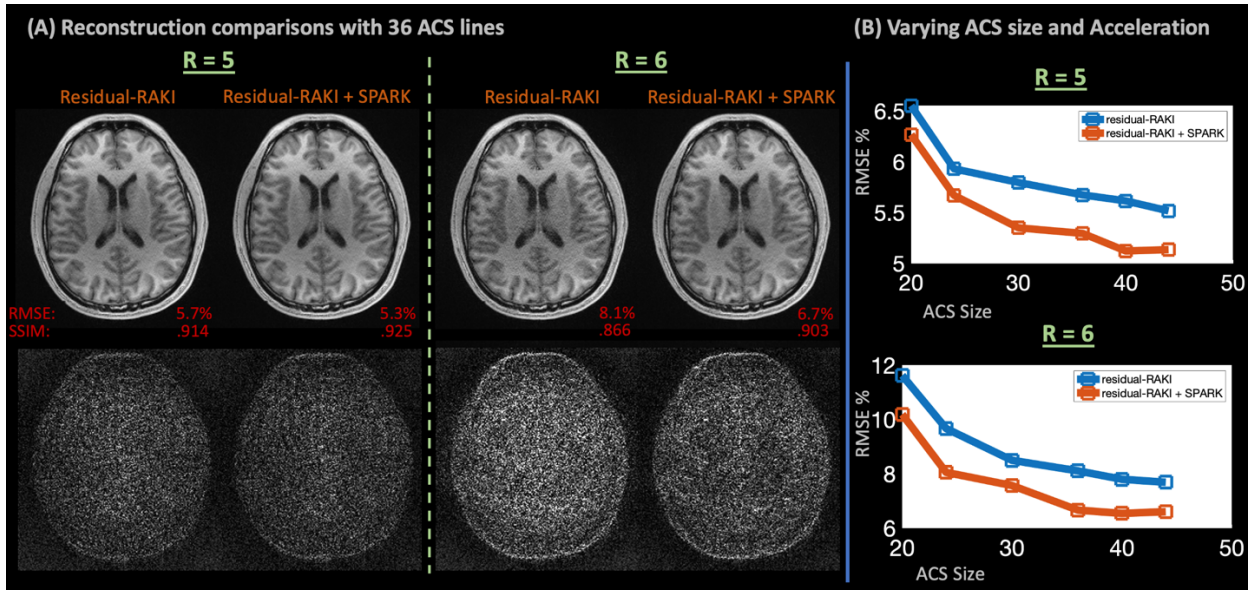

*Fig 4.* (A) Applying SPARK to residual-RAKI with 36 ACS lines and $R_{in-plane}$ = {5,6}. Combining residual-RAKI with SPARK reduces RMSE, SSIM, and error. (B) RMSE values for residual-RAKI and residual-RAKI with SPARK at $R_{in-plane}$ = {5,6} and a range of ACS sizes. Combining SPARK with residual-RAKI improves performance with all ACS sizes and acceleration factors suggesting that SPARK and residual-RAKI can be synergistically combined to improve reconstruction quality.

**Fig 5** compares GRAPPA and GRAPPA with SPARK using our proxy for retained SNR. Averaging across the brain, GRAPPA achieves values of {24.9,15.8,9.0}, while GRAPPA with SPARK achieves {27.0,20.6,16.0} for $R_{in-plane}$ = {4,5,6} respectively. Across the 100 instances, GRAPPA with SPARK improves upon the mean reconstruction RMSE and SSIM of GRAPPA with similar variance for $R_{in-plane}$ = {4,5,6}. This suggests that SPARK improves our proxy for retained signal while improving image quality.

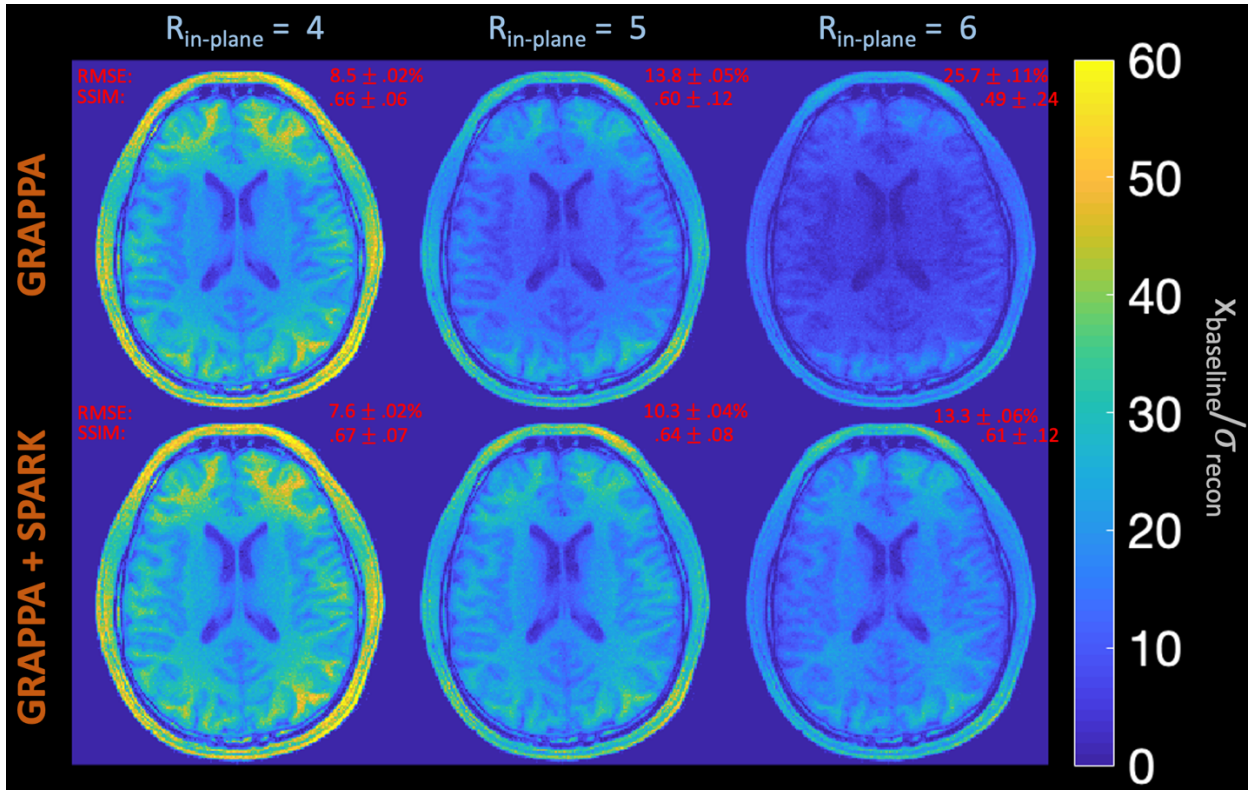

*Fig 5.* Comparing our proxy for retained SNR between GRAPPA and GRAPPA with SPARK reconstructions computed using a version of the pseudo-replica technique with synthesized k-spaces from 100 different noise instances for $R_{in\text{-}plane}$ = {4,5,6}. At all acceleration rates, GRAPPA with SPARK retains equivalent or higher values in comparison to GRAPPA alone. For $R_{in\text{-}plane}$ = {5,6}, adding SPARK refinement yields significant improvements particularly around the center of the brain.

**Supporting Fig 1** visualizes the SPARK k-space correction in the individual coil and coil-combined k-spaces for the $R_{in\text{-}plane}$ = 6 baseline GRAPPA reconstruction used to compute our proxy for retained SNR.

**Fig 6** illustrates how SPARK synergizes with advanced parallel imaging techniques like VC-GRAPPA and LORAKS. At $R_{in\text{-}plane}$ = {5,6}-fold acceleration with 30 ACS lines, VC-GRAPPA and LORAKS improve RMSE And SSIM in comparison to standard GRAPPA. SPARK further refines these input techniques, both visually, as seen by the error images, and quantitatively with improved RMSE and SSIM. Combining SPARK with LORAKS and VC GRAPPA also outperforms residual-RAKI in quantitative metrics and residual error maps.

**Supporting Fig 4** presents comparisons between VC-GRAPPA, LORAKS, VC-GRAPPA with SPARK, and LORAKS with SPARK for $R_{in\text{-}plane}$ = {4,7}. Applying SPARK to VC-GRAPPA and LORAKS improves reconstruction RMSE and SSIM at both the modest and large acceleration rates.

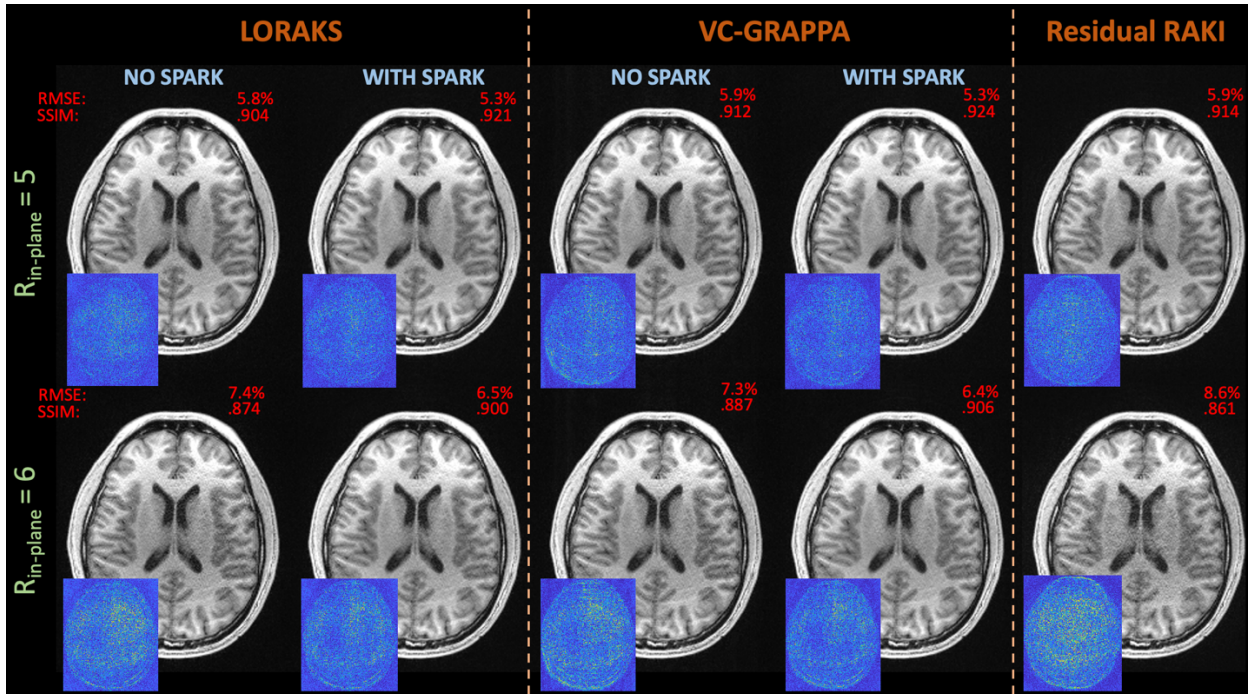

*Fig 6.* SPARK applied to two advanced parallel imaging techniques, LORAKS and VC-GRAPPA, and residual-RAKI in a slice of the cartesian MPRAGE dataset. While LORAKS, VC-GRAPPA, and residual-RAKI produce reasonable images, synergistically applying SPARK with LORAKS and VC-GRAPPA input yields further refinement through RMSE and SSIM improvement and visual error reduction.

**Supporting Fig 5** compares SPARK applied to GRAPPA and VC-GRAPPA at $R_{in-plane}$ = {5,6}. At both acceleration rates, VC-GRAPPA outperforms GRAPPA. For $R_{in-plane}$ = 5, applying SPARK to both VC-GRAPPA and GRAPPA improves quantitative metrics and yields similar quality images. $R_{in-plane}$ = 6 reconstructions yield analagous results, but VC-GRAPPA with SPARK slightly outperforms GRAPPA with SPARK in RMSE and SSIM.

**Supporting Fig 6** displays GRAPPA reconstructions with Tikhonov regularization values [0, 1, 5, 10, 50] and the associated SPARK correction at $R_{in-plane}$ = 5. At all regularization values, applying SPARK to GRAPPA improves RMSE, SSIM, and qualitative image quality.

**Supporting Fig 7** showcases SPARK corrections applied to GRAPPA and LORAKS reconstructions at $R_{in-plane}$ = {6,7} on the high SNR slice. At both acceleration rates, LORAKS outperforms GRAPPA, but, applying SPARK to either technique significantly improves RMSE, SSIM, and qualitative image quality.

In-vivo Volumetric Cartesian Imaging Results

**Fig 7** shows representative axial, coronal, and sagittal slices comparing SPARK to GRAPPA in 3D. Applying SPARK to GRAPPA significantly improves RMSE and SSIM, both volumetrically and in representative slices. Additionally, SPARK achieves qualitative and quantitative improvement in the three representative slices without significant blurring artifacts.

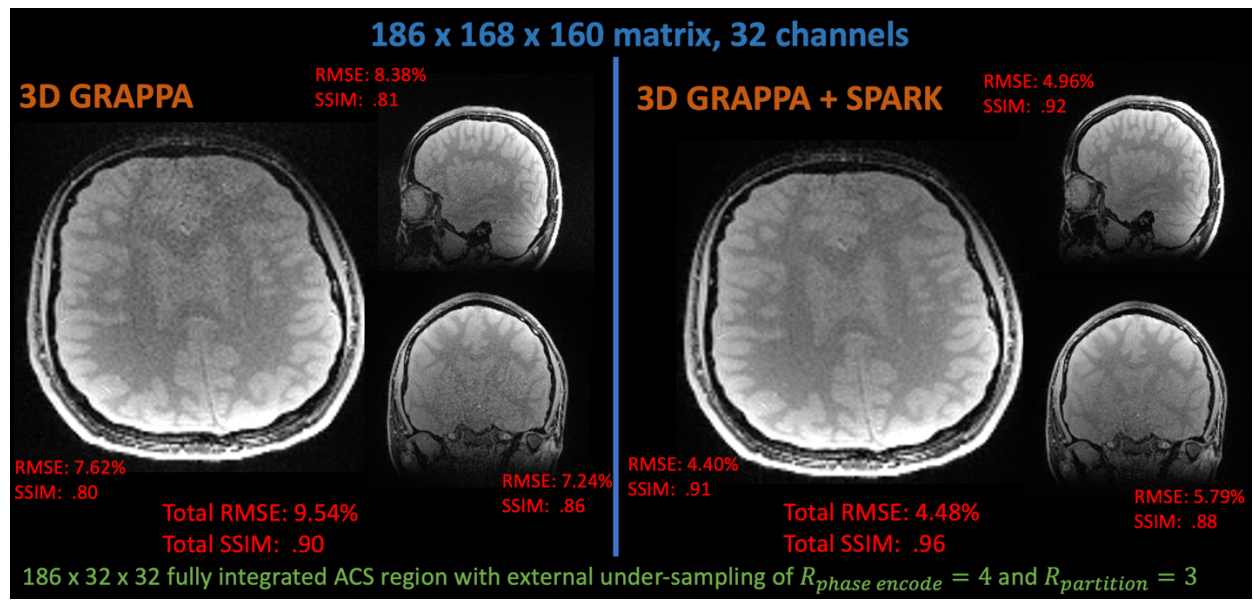

*Fig 7.* 3D GRE comparisons between 3D GRAPPA and the associated 3D SPARK reconstruction for three representative slices with a 32 x 32 integrated ACS region and undersampling in the phase encode and partition dimensions by a factor of 4 and 3 respectively. SPARK achieves SSIM and 2x RMSE improvement over the whole volume and produces cleaner images illustrated by the axial, coronal, and sagittal slices displayed.

**Fig 8** depicts the schematic 12x net acceleration sampling patterns and representative slices of 3D GRAPPA and 3D GRAPPA with SPARK reconstructed without a measured ACS region and without regularization in the GRAPPA reconstructions. SPARK achieves RMSE and SSIM improvement both in the entire volume as well as the representative slices. Spark also demonstrates qualitative visual reduction of noise amplification and aliasing artifacts seen in the 3D GRAPPA reconstruction.

**Supporting Figure 8** reports the same comparisons from **Fig 8**, but with minimum RMSE Tikhonov regularization used in the GRAPPA input to SPARK and the baseline 3D GRAPPA reconstruction. SPARK achieves less RMSE and SSIM improvement (in comparison to Figure 8) since the regularization parameter in 3D GRAPPA minimizes RMSE. However, this over regularization comes at the cost of structured aliasing artifacts, SPARK yields improved quantitative metrics and produces cleaner images without the artifacts present in the GRAPPA images.

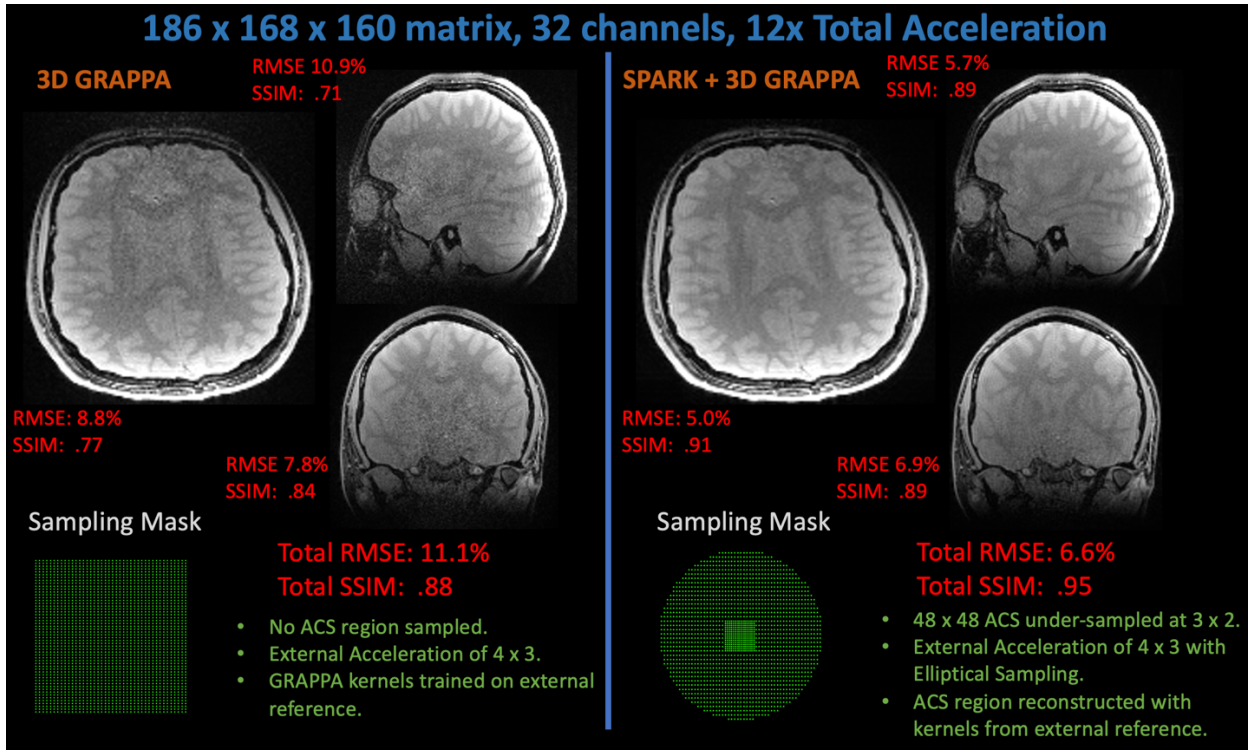

*Fig 8.* Comparisons between 3D GRAPPA using kernels trained on a reference scan and acceleration matched SPARK with GRAPPA-input. SPARK used k-space data with a 48 x 48 ACS region with 3 x 2 undersampling and exterior of k-space with 4 x 3 undersampling, while GRAPPA reconstructed 4 x 3 uniformly undersampled data. SPARK achieves significant RMSE and SSIM improvement over the whole volume and produces cleaner images illustrated by the axial, coronal, and sagittal slices displayed.

In-vivo Wave-encoded Imaging Results

In **Fig 9**, SPARK compares wave-encoding and wave-encoding with SPARK reconstructions for the wave-encoded MPRAGE slice at acceleration factors of {5,6}. Applying SPARK to the wave-encoded reconstruction improves RMSE, SSIM, and qualitative image quality.

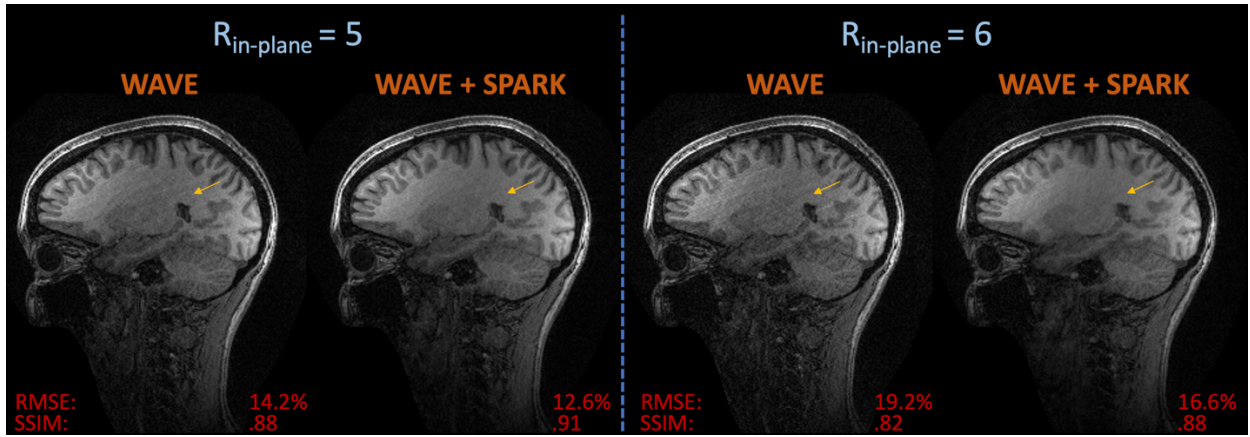

*Fig 9.* Wave-encoding reconstructions with and without SPARK of the MPRAGE dataset for $R_{in-plane}$ = {5,6} and 24 ACS lines.  SPARK flexibly synergizes with the wave-encoded paradigm as wave-encoding with SPARK improves RMSE and SSIM values and produces qualitatively cleaner images in comparison to wave-encoding alone.

**Supporting Fig 2** shows the individual coil and coil-combined k-spaces for the wave-encoded reconstruction, estimated SPARK correction, and the corrected k-space for the experiment with an acceleration factor of 6.

**Fig 10** compares standard wave-encoding and wave-encoding with SPARK on the MPRAGE slice group.  Wave-encoding with SPARK improves RMSE, SSIM, and visual results, by combining improved conditioning from wave-encoding and SPARK corrections.

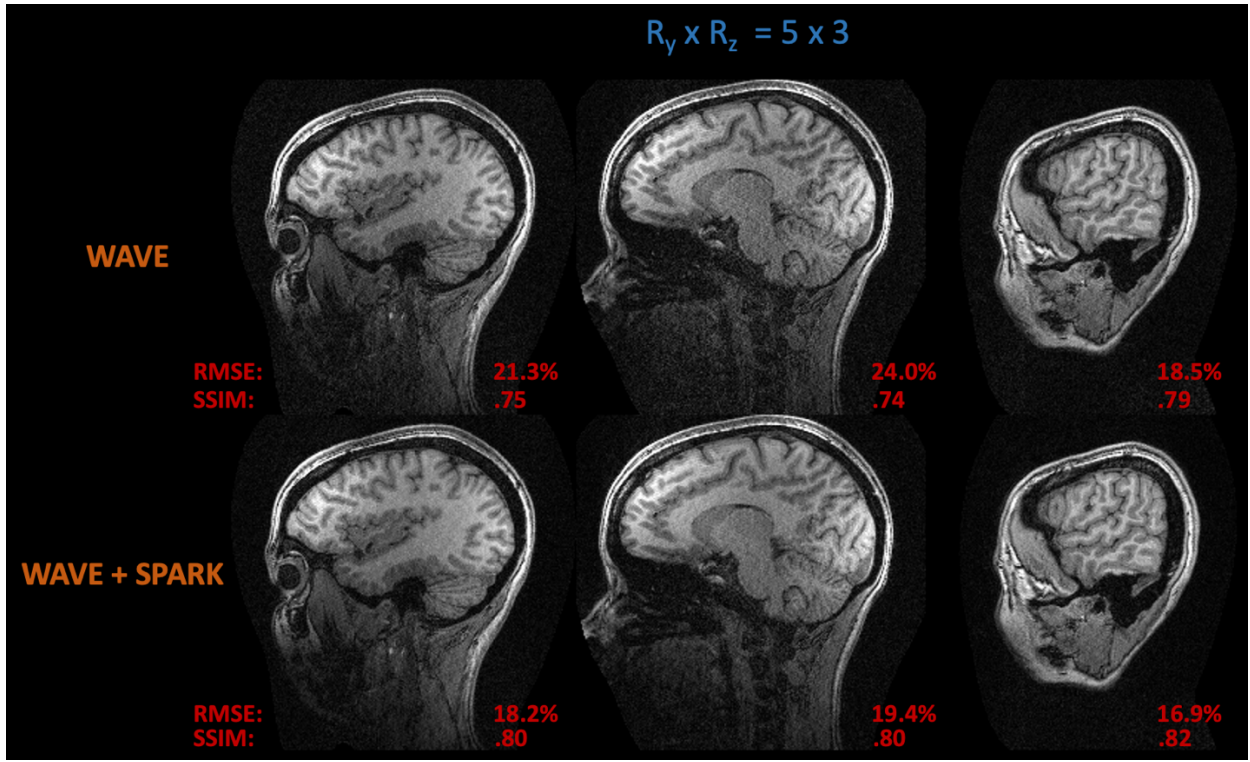

*Fig 10.* Wave-encoding reconstructions with and without SPARK of the wave-encoded MPRAGE slice-group for $R_{phase-encode}$ = 5, $R_{partition}$ = 3. Applying SPARK to the wave-encoded reconstruction improves quantitative and qualitative results in comparison to wave-encoding alone, demonstrating that SPARK synergizes with non-cartesian, 3D wave-encoded acquisitions.

## DISCUSSION

SPARK extends the elegant idea of using ACS regions, introduced by RAKI, to train scan-specific models for accelerated MRI by correcting k-space artifacts of an arbitrary input reconstruction. SPARK with GRAPPA input exhibits improved performance over GRAPPA and robustness to ACS size over RAKI and residual-RAKI. Moreover, we found that SPARK synergizes with residual-RAKI. We also demonstrated how SPARK applies to several acquisition and reconstruction scenarios like VC-GRAPPA and LORAKS, 3D reconstructions without a fully sampled ACS region, and non-cartesian wave encoding imaging.

SPARK operates in k-space as the ACS region enables learning without access to a database of training samples. In scan-specific settings, performing image-based error correction is less feasible. Although the ACS region corresponds to a fully sampled low-resolution image, image-based errors associated with a low-resolution reconstruction do not correspond to the image-based reconstruction artifacts present in the higher resolution accelerated MRI reconstruction. Image based residual estimation works well when a database of accelerated reconstructions and associated fully sampled images exists[29–32], but k-space based techniques are more conducive for error correction in a scan-specific setting.

Although SPARK bears resemblance to residual-RAKI[38], several differences persist. Residual-RAKI trains two scan-specific networks to simultaneously perform interpolation of missing k-space and correction of k-space interpolation artifacts. In addition, the interpolation network is constrained to perform a linear reconstruction on 2D cartesian acquisitions. SPARK, however, trains a single scan-specific network to estimate reconstruction artifacts given an input reconstruction unconstrained to linear convolutions on 2D cartesian acquisitions. As a result, SPARK, to the best of our knowledge, is the first scan-specific k-space correction technique applied to advanced parallel imaging techniques like VC-GRAPPA, 3D acquisitions without an ACS region, and wave-encoded acquisitions. Finally, SPARK and residual-RAKI can be complementary techniques, as we successfully applied SPARK to residual-RAKI to improve the reconstruction quality relative to either residual-RAKI or SPARK with GRAPPA.

Figure 3 and Supporting Figure 3 suggest that SPARK applied to GRAPPA outperforms RAKI and residual-RAKI in acquisitions with smaller ACS regions. Since RAKI and residual-RAKI directly interpolate missing k-space data, only the ACS region is utilized to train the scan-specific networks. In addition, RAKI and residual-RAKI apply three consecutive convolutional kernels that truncate the spatial dimensions of the network input. After taking the input ACS through a CNN for training, the truncated output can only be compared to a subset of the total ACS region for backpropagation. Since SPARK takes reconstructed total k-space as input, backpropagation can be performed by comparing all the measured and reconstructed ACS region. This increased efficiency in ACS usage during training may explain SPARK's improved performance at smaller ACS sizes.

We envision SPARK providing clinical utility in two regimes (amongst others). First, RAKI (and subsequently residual-RAKI) were proposed to improve reconstructions when it is not feasible to acquire fully sampled data; settings with extensive motion, like fetal, cardiac, and abdominal MRI, dynamic MRI, and time-resolved imaging. SPARK could provide utility in these scenarios since it functions in a scan-specific fashion as well. Second, several recent publications describe the utility of 3D wave-encoding in clinical applications[60–63]. However, images still suffer from increased noise levels, and applying SPARK might further push the limits of 3D-wave-encoded applications by enabling higher acceleration rates or improving image quality.

Figure 6, Supporting Figure 4, and Figure 9 show experiments in which SPARK synergizes with advanced 2D parallel imaging techniques and wave-encoding. At acceleration rates of 6 and 7, the resultant cartesian and wave-encoded SPARK corrected and residual-RAKI images exhibit noticeable improvement but may suffer from residual-aliasing artifacts. For $R_{in-plane}$ = 5, the images are similar, with subtle differences captured by the error images. The proof-of-concept experiments demonstrate that SPARK can synergize with advanced techniques, but future work requires further study of acceleration factors, imaging settings, and reconstruction algorithm combinations.

Figures 3, 4, 6 show that combining SPARK with an advanced technique like residual-RAKI, LORAKS, or VC-GRAPPA improves performance in comparison to applying SPARK to GRAPPA. Since SPARK aims to estimate

and correct reconstruction artifacts, we hypothesize that in scan-specific settings, estimating errors from an improved reconstruction might be an easier problem than estimating errors from a relatively worse reconstruction.

In Figure 8, the more densely sampled ACS may contribute to performance difference between SPARK with GRAPPA and uniformly under-sampled GRAPPA. However, we aimed to provide a proof-of-principle application in which SPARK can be applied to GRAPPA without a fully sampled ACS region in comparison to an acceleration matched GRAPPA reconstruction. In addition, Figure 7 demonstrates that SPARK provides significant improvement in comparison to GRAPPA with internal ACS. SPARK also improves the GRAPPA input to SPARK used in Figure 8 (with the same, more densely sampled ACS region) in a similar fashion to the uniformly under-sampled GRAPPA reconstruction.

While the presented MPRAGE experiments employed retrospective under-sampling, prospective implementations may be affected by the interplay between signal evolution and sampling within each TR. However, prospective sampling schemes can be developed that minimize these adverse effects.[64]

*Limitations*

SPARK assumes k-space reconstruction errors behave similarly inside and outside of the ACS region. As a result, SPARK effectively corrects noise amplification, but less significantly improves obvious and local structured aliasing artifacts at extreme acceleration rates. High acceleration requires combining SPARK with a more advanced technique, like the wave-encoding studies in Figures 9 and 10, to reduce structured aliasing and suppress noise amplification.

Computation for the presented 3D cartesian experiments in Figures 7 and 8 can be lengthy (2 – 3 days) because SPARK operates on the entire volumetric k-space simultaneously, rather than performing an inverse-Fourier transform along the readout-dimension and operating on individual slices. We choose these experiments to demonstrate proof-of-concept novelty in training on a 3D slab of ACS and applying corrections to the 3D volume to improve reconstruction performance. However, computation for our fully volumetric 3D experiments can be significantly reduced by lowering the number of iterations for training and applying geometric coil compression[65].

**Supporting Figure 11** compares 3D-GRAPPA to SPARK + GRAPPA with 200 training iterations and 16-channel coil-compressed k-space on the 3D GRE dataset from Figures 7 and 8 with 3 x 3 under-sampling. SPARK + GRAPPA achieves RMSE and SSIM improvement with total computation time reduced from 2 days to just 62 minutes. A parallelized implementation that trains models for all coils simultaneously might further reduce computation time by a factor of 16. We have also recently developed a scan-specific technique, which could be combined with SPARK, that exploits explicit knowledge of the coil sensitivity profiles to drastically reduce training times[66].

If the learning rate and optimizer pairing is chosen inappropriately, we have seen SPARK reduce image quality. This limitation is not unique to SPARK, as many CNN based approaches like RAKI and other standard supervised networks require careful tuning of hyper-parameters.[34,42,67]

## CONCLUSIONS

We introduce a scan-specific model that improves accelerated MRI by correcting k-space artifacts in a range of advanced reconstruction inputs. We apply SPARK to 2D cartesian GRAPPA, LORAKS, and VC-GRAPPA, residual-RAKI, 3D cartesian GRAPPA with and without ACS regions, and non-cartesian, wave-encoded imaging.

## DATA AVAILABILITY STATEMENT

In the spirit of reproducible research, code and data used to generate **Fig 3 – 10** can be found at the following links respectively:

https://github.com/YaminArefeen/spark_mrm_2021
https://www.dropbox.com/sh/zveq2tfh7mgr9qk/AABSuSM23QOFVAe0SJ9oBIm6a?dl=0

## Acknowledgements

The authors thank the anonymous referees for their careful comments and feedback which significantly improved the quality of the manuscript.

This work was supported in part by research grants R01 EB017337, R01 HD100009, R03 EB031175, U01 EB025162, U01 EB026996 and P41 EB030006. In addition, this material is based upon work supported by the National Science Foundation Graduate Research Fellowship Program under Grant No. 1122374. Any opinions, findings, and conclusions or recommendations expressed in this material are those of the authors and do not necessarily reflect the views of the National Science Foundation.

Supporting Figure S1

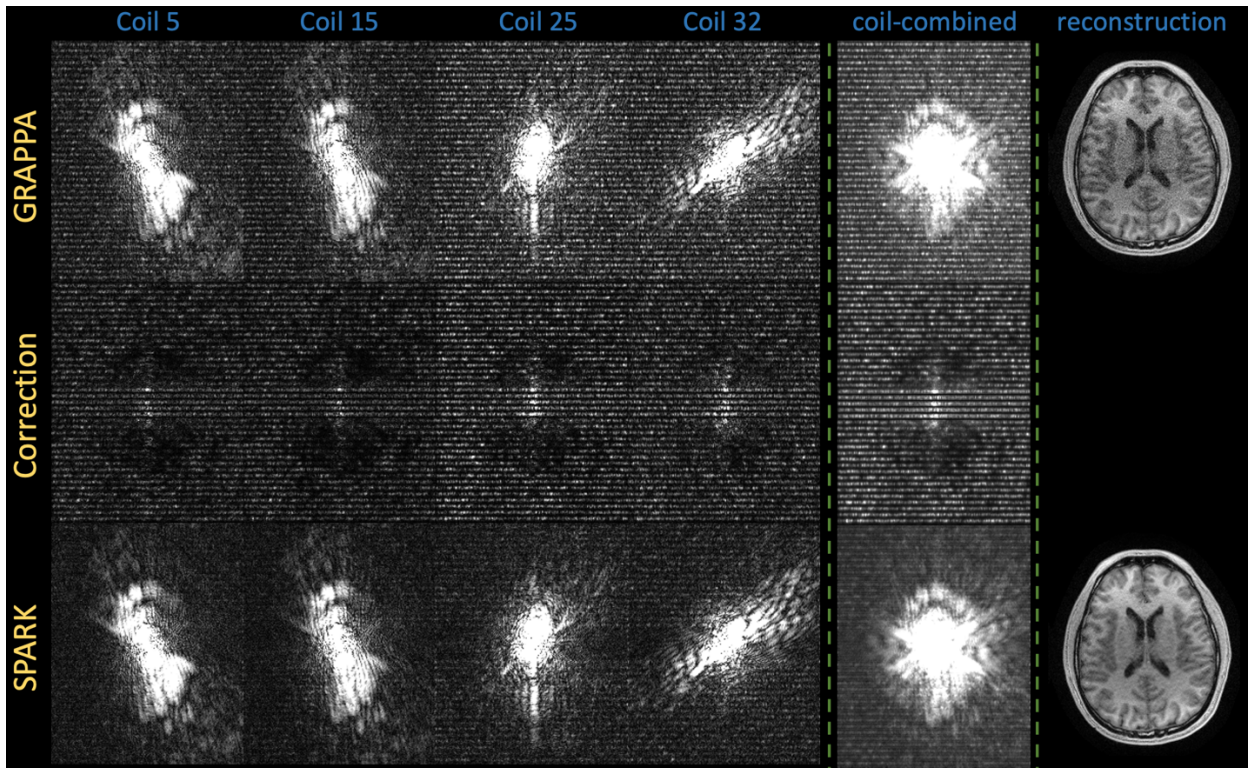

*Supporting Information Fig S1.* Individual coil and coil-combined k-spaces for the GRAPPA reconstruction, the estimated SPARK correction, and the corrected SPARK k-space for the MPRAGE slice undersampled with $R_{in-plane}$ = 6. Note, all k-space images are scaled by a factor of 100 for visualization. The SPARK model estimates correction terms which reduce the GRAPPA reconstruction errors in k-space, resulting in a cleaner reconstructed image.

Supporting Figure S2

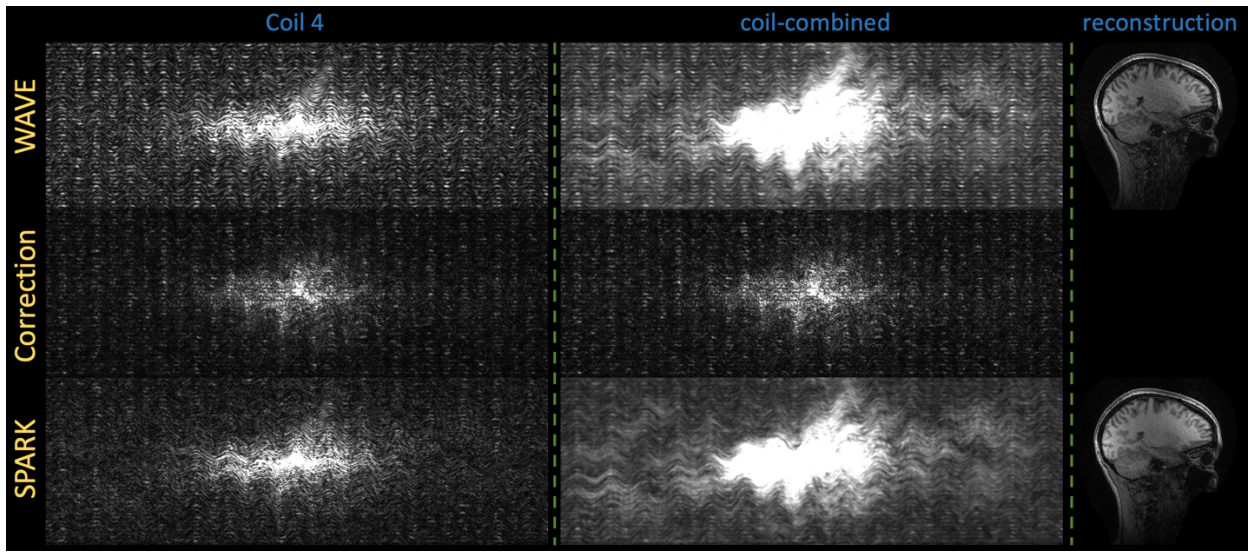

*Supporting Information Fig S2*. Individual coil and coil-combined k-spaces for the single-slice wave-encoded reconstruction, the estimated SPARK correction, and the corrected SPARK k-space for the wave-encoded dataset undersampled with $R_{in-plane}$ = 6. Note, all k-space images are scaled by a factor of 100 for visualization. The SPARK model estimates correction terms which reduce the generalized-SENSE reconstruction errors in k-space, resulting in a cleaner reconstructed image.

Supporting Figure S3

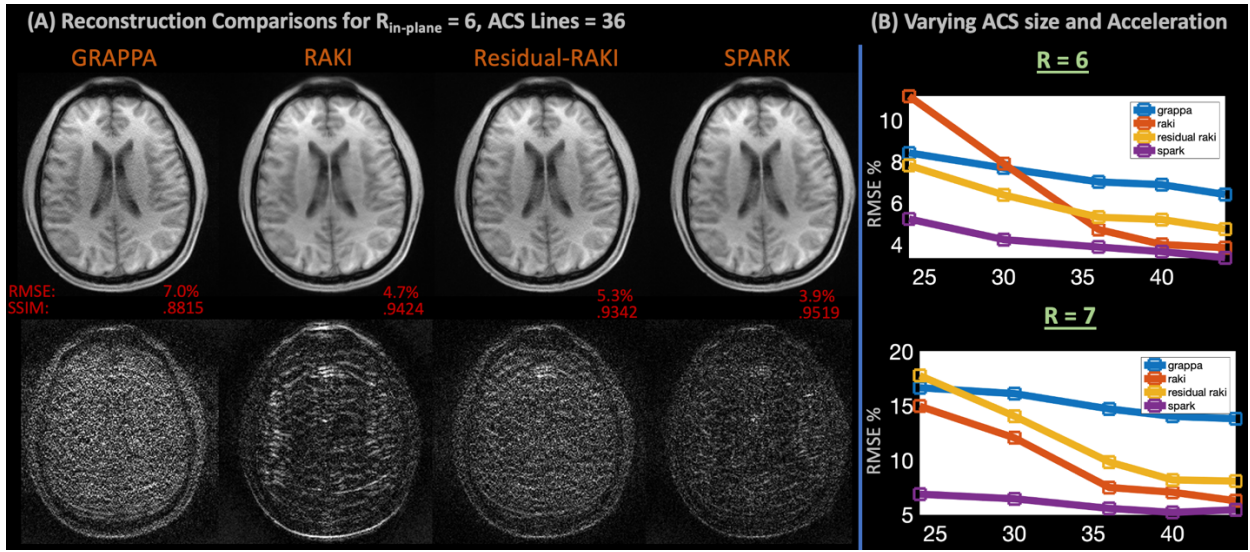

*Supporting Information Fig S3.* A series of comparisons between GRAPPA, RAKI, residual-RAKI, and SPARK on the high SNR, MPRAGE slice.  *(A)* Images of GRAPPA, RAKI, residual-RAKI, and SPARK applied to an initial GRAPPA reconstruction at $R_{in\text{-}plane}$ = 6 and 36 ACS lines.  Note, RAKI reduces RMSE in comparison to residual-RAKI in some cases on this high SNR dataset.  Looking at the error maps in (A), it seems like RAKI yields less noise amplification but more structured residual aliasing artifacts.  As a result, the L2-error metric of RMSE captures the improvement in noise-amplification but may struggle with quantifying errors from structured artifacts.  This might not be true in low SNR settings due to the relative change in noise efficiency of the two techniques. RAKI, residual-RAKI, and SPARK all achieve quantitative RMSE, SSIM, and visual improvements in comparison to GRAPPA.  *(B)* Comparisons between GRAPPA, RAKI, residual-RAKI, and SPARK for $R_{in\text{-}plane}$ = {6,7} and a range of ACS sizes. SPARK always performs at least as well or better than RAKI and residual-RAKI, particularly at smaller ACS sizes.

Supporting Figure S4

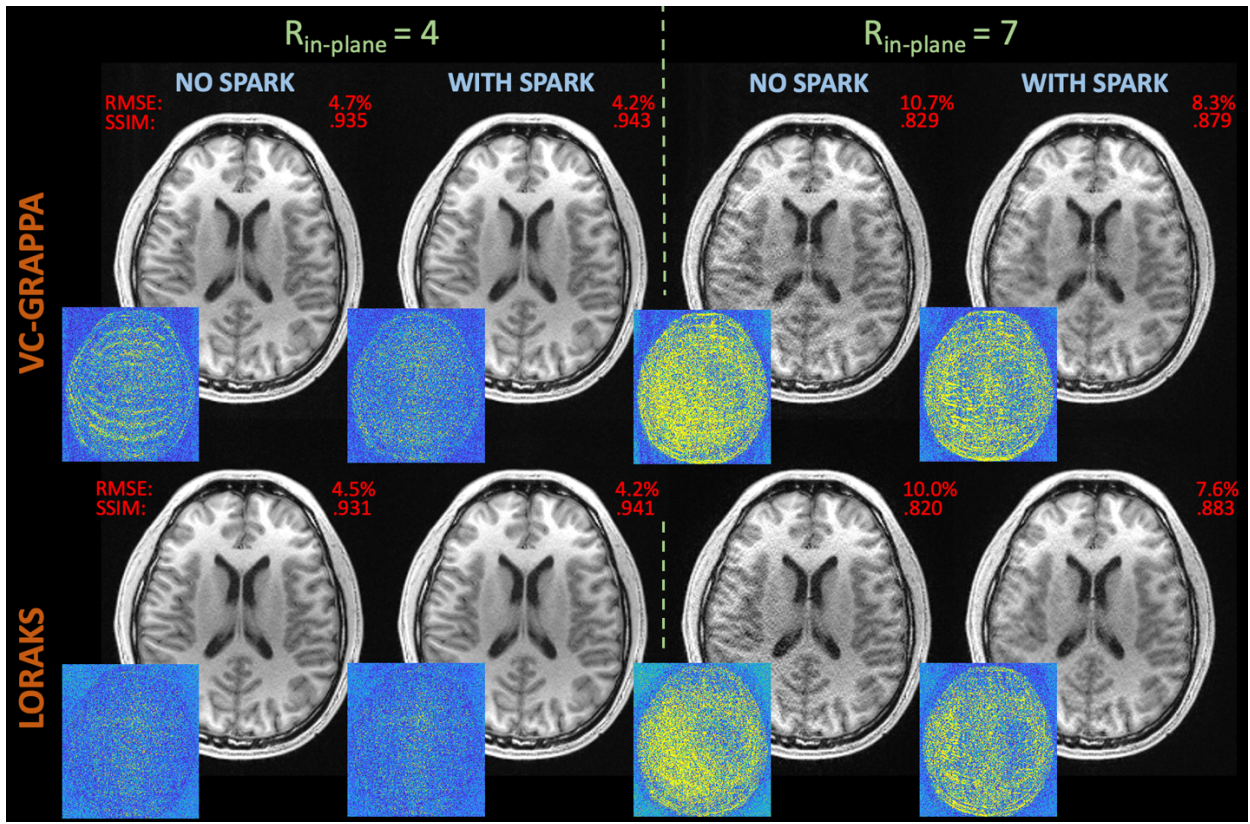

***Supporting Information Fig S4***. SPARK applied to VC-GRAPPA and LORAKS on the MPRAGE slice with $R_{in\text{-}plane}$ = {4,7}. At an acceleration factor of 4, SPARK yields modest RMSE, SSIM, and visual improvement and yields relatively clean images. At an extreme acceleration factor of 7, image quality degrades, but SPARK provides substantial RMSE, SSIM, and qualitative improvement for both VC-GRAPPA and LORAKS.

Supporting Figure S5

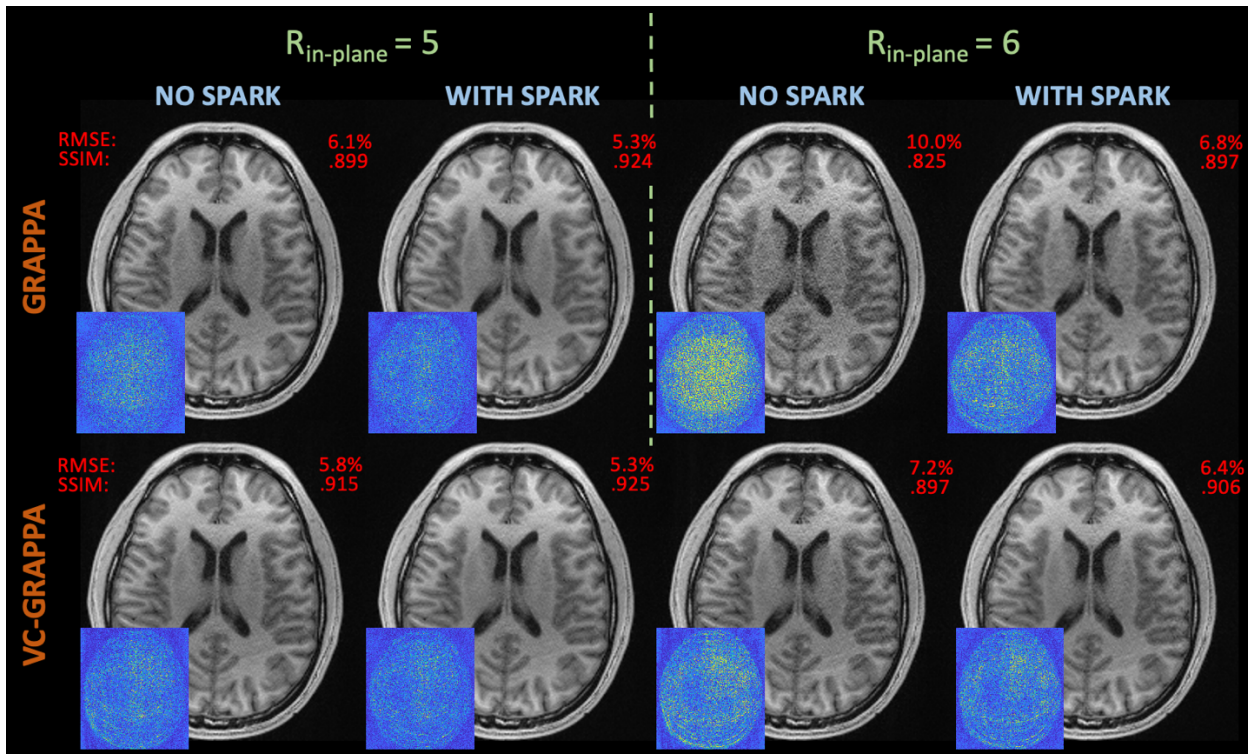

***Supporting Information Fig S5***. SPARK applied to GRAPPA and VC-GRAPPA at $R_{in\text{-}plane}$ = {5,6}. VC-GRAPPA achieves lower RMSE by reducing noise amplification, particularly around the center of the brain. After SPARK correction at $R_{in\text{-}plane}$ = 5, both techniques yield very similar RMSE, SSIM, and images, suggesting that SPARK brings images with different noise-levels but similar artifacts closer after correction. Larger differences between the SPARK reconstructions persist at $R_{in\text{-}plane}$ = 6, but the images with SPARK are still more similar than the corresponding GRAPPA and VC-GRAPPA reconstructions without SPARK.

Supporting Figure S6

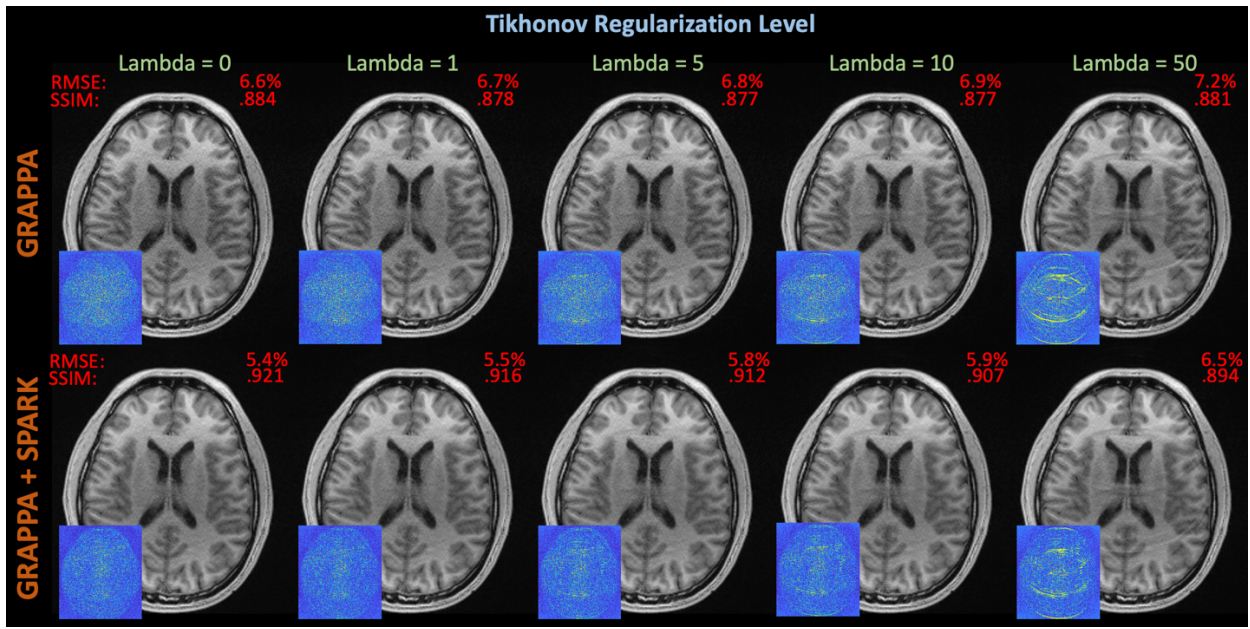

***Supporting Information Fig S6.*** SPARK applied to GRAPPA reconstructions with Tikhonov regularization values [0, 1, 5, 10, 60] at $R_{in\text{-}plane}$ = 5. At all regularization levels, SPARK improves GRAPPA RMSE and SSIM and reduces visual artifacts. At lower Tikhonov regularization values, like [0, 1], SPARK predominantly corrects noise amplification artifacts. However, SPARK reduces both structured aliasing and noise amplification artifacts when applied to the overly regularized GRAPPA reconstructions.

Supporting Figure S7

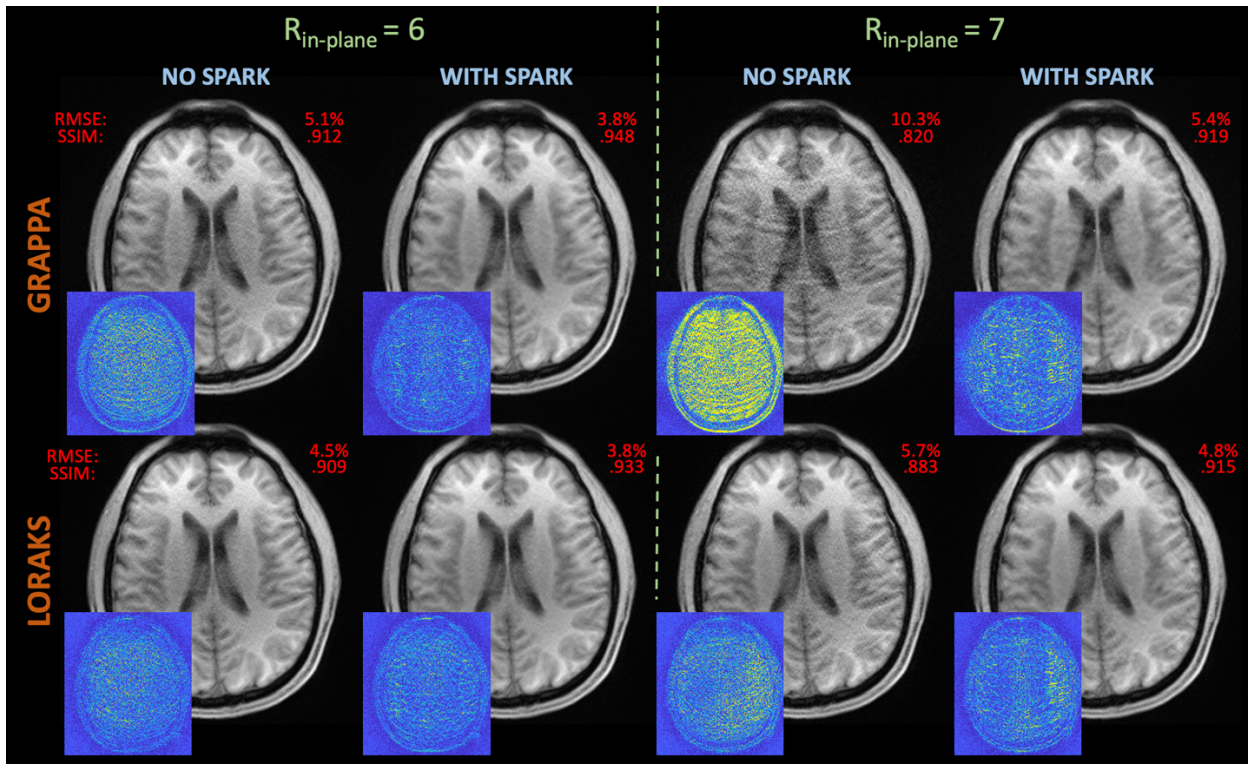

***Supporting Information Fig S7.*** SPARK applied to GRAPPA and LORAKS reconstructions of the high SNR 1 x 1 x 10 mm MPRAGE slice with $R_{in\text{-}plane}$ = {6,7} under-sampling. Even with higher SNR data, applying SPARK to both GRAPPA and LORAKS improves reconstruction performance at both acceleration rates, as evidenced by the RMSE and SSIM values and visual error reduction.

Supporting Figure S8

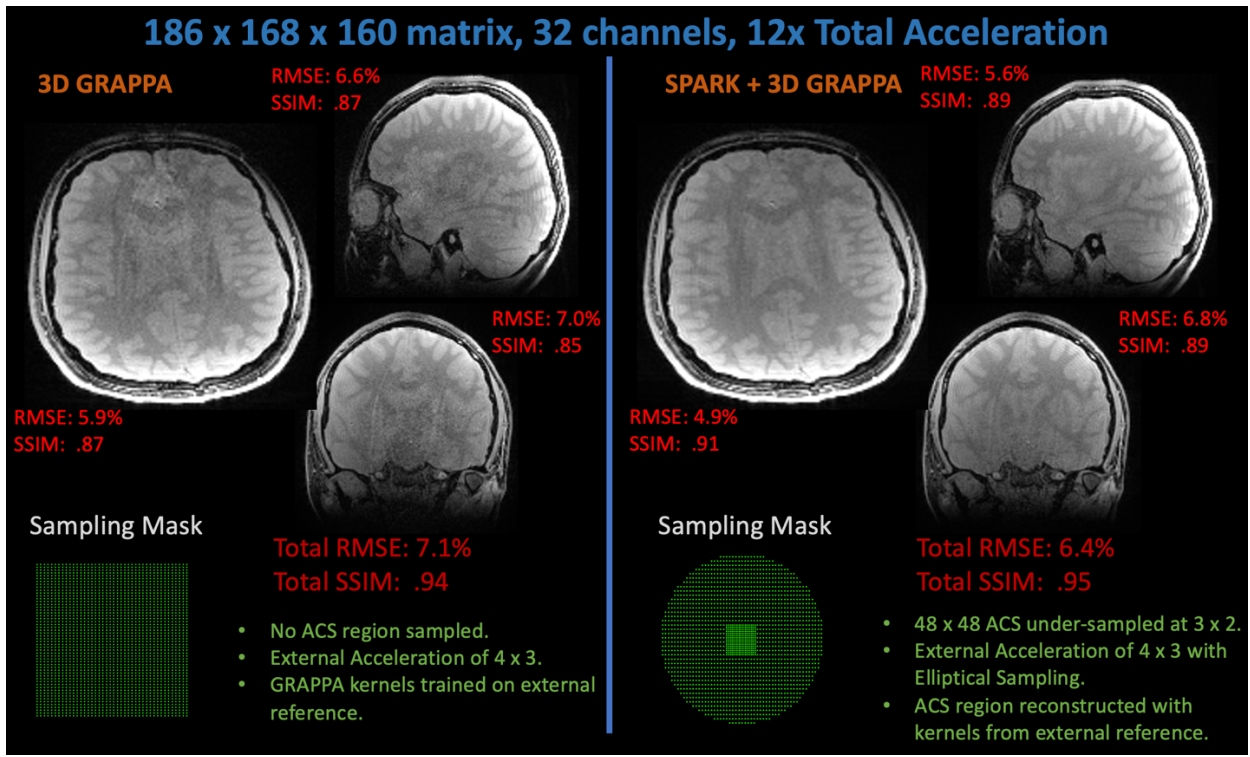

*Supporting Information Fig S8.* Comparisons between regularized 3D GRAPPA using 3D kernels trained on a reference scan and acceleration matched SPARK with regularized GRAPPA-input. SPARK used k-space data with a 48 x 48 ACS region with 3 x 2 undersampling and exterior of k-space with 4 x 3 undersampling, while GRAPPA reconstructed 4 x 3 uniformly undersampled data. While both techniques yield similar total RMSE and SSIM, GRAPPA suffers from structured biases in the representative slices, induced by the Tikhonov regularization parameter. SPARK improves RMSE and SSIM metrics and yields qualitatively cleaner images by reducing structured aliasing artifacts produced by regularization in comparison to GRAPPA.

Supporting Figure S9

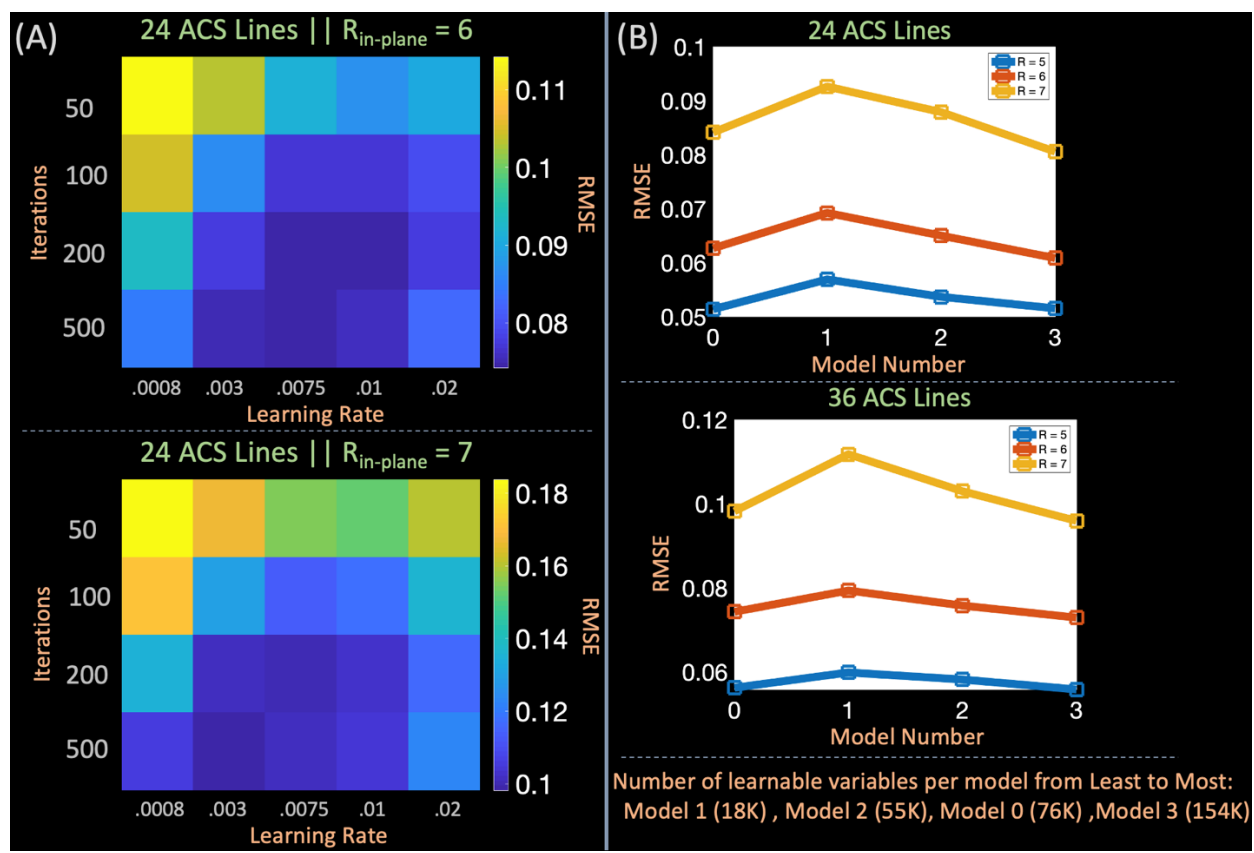

***Supporting Information Fig S9.*** (A) RMSE values of SPARK + GRAPPA when applying the 2D model from Figure 2 (a) at $R_{in-plane}$ = {6,7} with 24 calibration lines and varying iterations and learning rates. At $R_{in-plane}$ = 6, 200 iterations with a learning rate of .0075 and .01 yield the lowest RMSE, while learning rate and iteration pairs .003 / 500 and .0075 / 200 yield the lowest RMSE at $R_{in-plane}$ = 7. In order to limit computation time and safeguard against potential instability from large learning rates, we choose a learning rate of .0075 with 200 iterations for all experiments that utilize our 2D model from Figure 2 (a). (B) Performance of models that differ in complexity at $R_{in-plane}$ = {6,7} and 24 or 36 ACS lines. Model 0 (76K trainable variables) is the 2D model used throughout the manuscript. Model 1 (18K trainable variables) and Model 2 (55K trainable variables) are less complex versions of Model 0. Model 3 (154K trainable variables) is a more complex version of Model 0. Model 0 and Model 3 comparably outperform Model 1 and Model 2. For the manuscript, we use Model 0 for all 2D experiments to achieve a reasonable tradeoff between performance, model complexity, and computation.

Supporting Figure S10

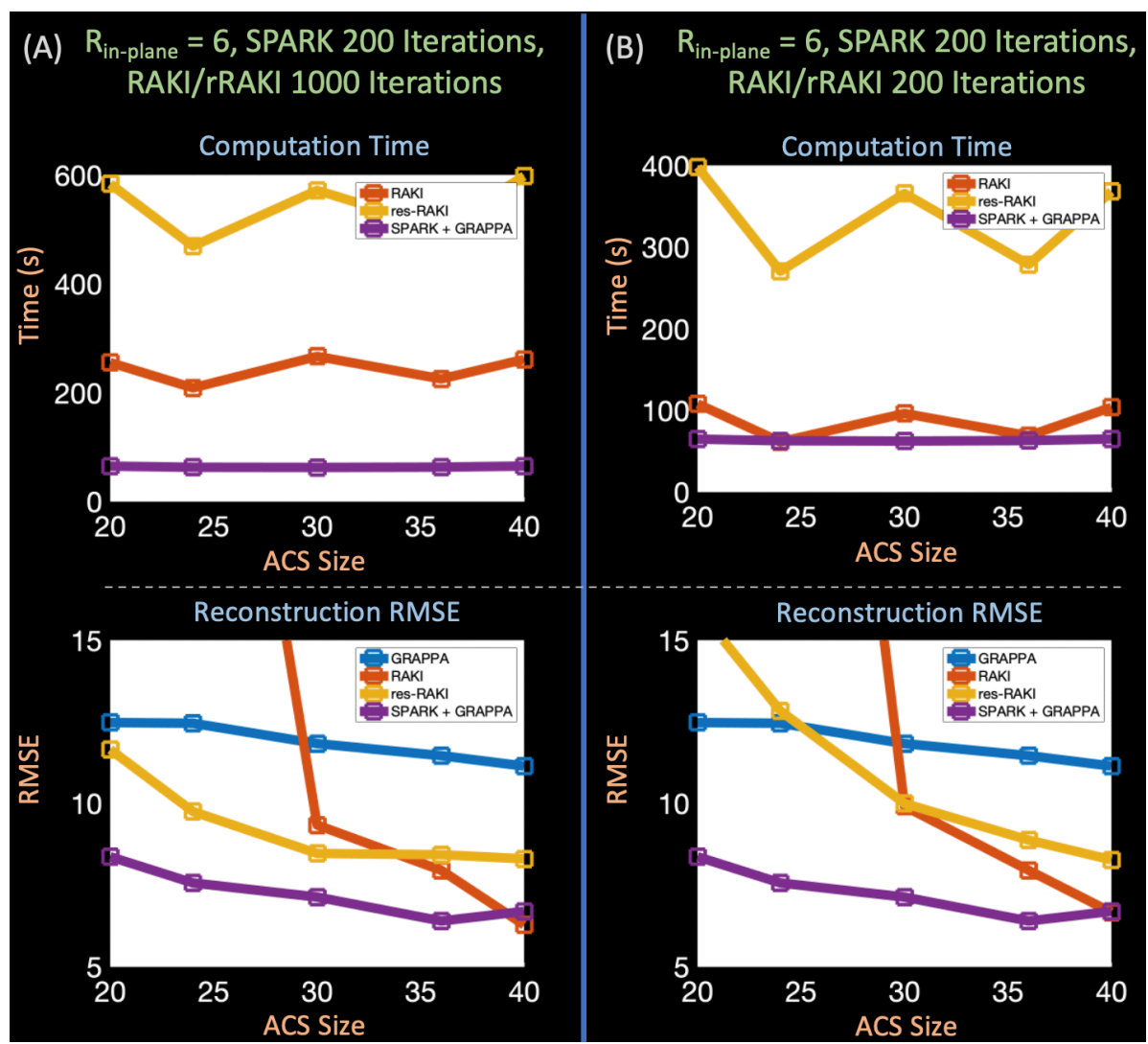

***Supporting Information Fig S10.*** (A) Total computation time and RMSE comparisons between SPARK + GRAPPA, RAKI, and residual-RAKI at $R_{in\text{-}plane}$ = 6 and varying ACS sizes. SPARK + GRAPPA models are trained with 200 iterations, and the RAKI / residual-RAKI models are trained with the default 1000 iterations. At all ACS sizes, SPARK + GRAPPA reduces computation and improves upon or matches RMSE in comparison to RAKI and residual-RAKI. (B) Total computation time and RMSE comparisons between SPARK + GRAPPA, RAKI, and residual-RAKI where models from all three techniques are trained with 200 iterations. SPARK + GRAPPA and RAKI achieve similar computation times and outperform residual-RAKI computation times at all ACS sizes. On the other hand, both RAKI and residual-RAKI suffer from worse reconstruction performance, as evidenced by the higher RMSE values at various ACS sizes, since the networks do not converge after 200 iterations. Note, computation time comparisons were computed on a Tesla V100 GPU with a PyTorch implementation of SPARK and the publicly available Tensorflow implementation of RAKI. Optimized and parallelized code could reduce and change the computation time comparisons presented.

Supporting Figure S11

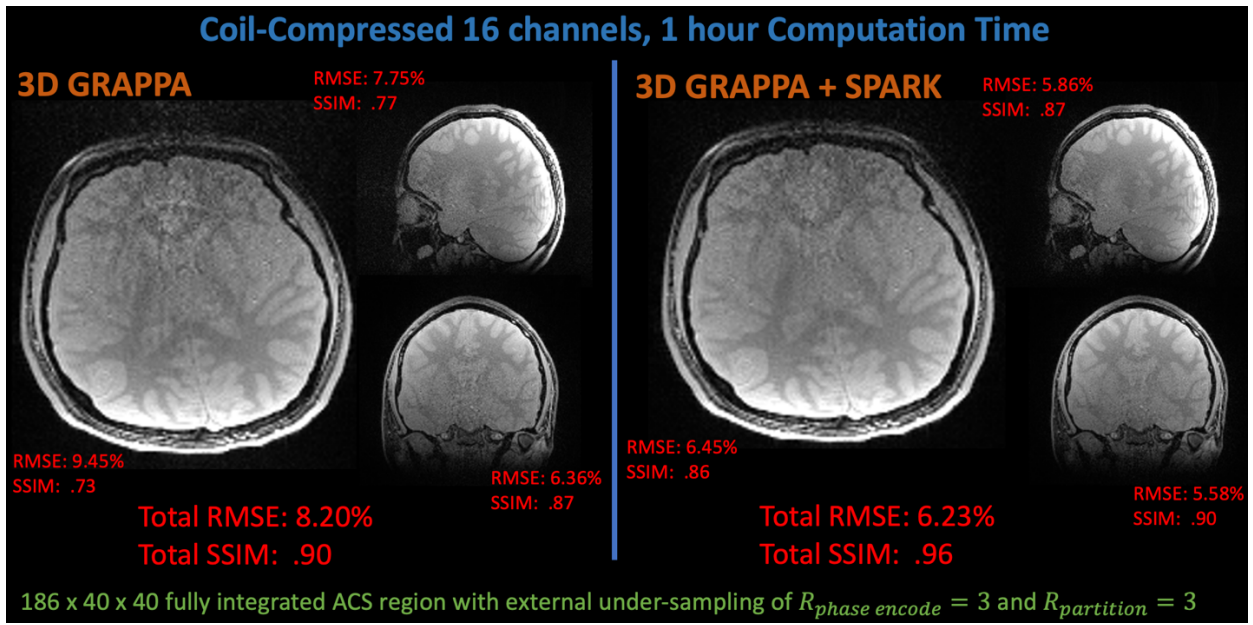

***Supporting Information Fig S11.*** 3D GRE comparisons between 3D GRAPPA and the associated 3D SPARK + GRAPPA reconstruction for three representative slices with a 40 x 40 integrated ACS region and undersampling in the phase encode and partition dimensions by a factor of 3 and 3 respectively. The data were compressed to 16 receive channels with geometric coil compression, and SPARK models were trained for 200 iterations. SPARK achieves RMSE and SSIM improvement over the whole volume and produces cleaner images illustrated by the axial, coronal, and sagittal slices displayed. With reduced training iterations and coil compression, computation time improved from 2 days to 62 minutes, a drastic decrease in comparison to the experiments presented in Figures 7 and 8.
38

# Supporting Text 1

Our chosen 3D model, illustrated in Figure 2 (b), consists of 391K trainable parameters. When applying SPARK to 3D GRAPPA, we found that less complex models (223K parameters or less than 150K parameters) did not estimate corrections as well as the 391K parameter model, evidenced by lower reconstruction RMSE throughout the volume. On the other hand, models with greater than 391K parameters did not fit into our GPU for training when applied to our 3D k-space dataset with a 186 x 168 x 160 matrix and 32 receive coils. Thus, we chose the 391K parameter model to achieve better reconstruction RMSE while meeting the memory requirements of our GPU.

The chosen 3D model was trained with 1000 iterations and a .002 learning rate. Experimentation revealed that higher learning rates sometimes caused instability in training, while lower learning rates required many more iterations for convergence. As a result, we found a learning rate of .002 to be a reasonable tradeoff between safeguarding against training instability and computation time. Additionally, 1000 iterations ensured that the training loss converged to some asymptote. However, we've found that similar quality results can be achieved with 200 iterations.